%% file: paper.tex
\documentclass[pdf,twocolumn,epjc]{svjour3}
\pdfoutput=1
\usepackage{multirow}
\usepackage{notoccite}
\usepackage{preprintcover}
\usepackage{amsmath}
\usepackage{graphicx}
\usepackage{afterpage}
\usepackage{overpic}
\usepackage{array}
\graphicspath{{}}
\usepackage{epstopdf}
\usepackage{subfigure}
\usepackage{placeins}
\usepackage{booktabs}
\usepackage{xspace}
\usepackage[switch]{lineno} 
\usepackage{cite}
\usepackage{hyperref}
\usepackage{siunitx}
\hypersetup{colorlinks,citecolor=blue,filecolor=blue,linkcolor=blue,urlcolor=blue}

\newcommand{\ph}{\phantom{x}}

\def\TeV{\ifmmode {\mathrm{\ Te\kern -0.1em V}}\else
                   \textrm{Te\kern -0.1em V}\fi}%
\def\GeV{\ifmmode {\mathrm{\ Ge\kern -0.1em V}}\else
                   \textrm{Ge\kern -0.1em V}\fi}%
\newcommand{\ttbar}{\ensuremath{t\bar{t}}}

\def\GeV{\ifmmode {\mathrm{\ Ge\kern -0.1em V}}\else
                   \textrm{Ge\kern -0.1em V}\fi}%

\input{Numbers.tex}

\def\thetitle{Search for dark matter in events with heavy quarks and missing transverse momentum in $pp$ collisions with the ATLAS detector}

\titlerunning{Search for DM with Heavy Quarks and \met\ with ATLAS}

\def\theabstract{%
This article reports on a search for dark matter pair production in association  with bottom or top quarks 
in \Lumi\ of $pp$ collisions collected  at  $\sqrt{s} = 8$~\TeV{} by the ATLAS detector at the LHC.
Events with large missing transverse momentum are selected when produced in association with 
high-momentum jets of which one or more are identified as jets containing $b$-quarks. Final states with top quarks are selected
by requiring a high jet multiplicity and in some cases a single lepton.
The data are found to be consistent with the Standard Model expectations and limits are set on 
the mass scale of effective field theories that describe scalar and tensor interactions between 
dark matter and Standard Model particles. Limits on the dark-matter--nucleon cross-section for 
spin-independent and spin-dependent interactions are also provided. These limits are particularly strong for low-mass dark matter. 
Using a simplified model, constraints are set on the mass of dark matter and of a coloured mediator suitable to explain a possible signal of annihilating dark matter.
}

\PreprintCoverPaperTitle{\thetitle}
\PreprintIdNumber{CERN-PH-EP-2014-229}
\PreprintCoverAbstract{\theabstract}
\PreprintJournalName{Eur.~Phys.~J.~C}

\begin{document}


\title{\thetitle}
\author{The ATLAS Collaboration}
\institute{CERN, 1211 Geneva 23, Switzerland, \email{atlas.publications@cern.ch}}
\date{Received:  16 October 2014 / Accepted: 9 February 2015} 
\journalname{Eur.~Phys.~J.~C} 

\maketitle

\abstract{\theabstract}

\section{Introduction}

The existence of dark matter (DM) in the Universe is highly motivated by many 
astrophysical and cosmological observations~\cite{Zwicky:1933gu, DMreview, Jungman:1995df, Binney:1993ce}. 
However, its nature remains a mystery. 
One of the best motivated candidates for a DM particle is a weakly interacting massive particle (WIMP)~\cite{Steigman:1984ac}.
At the Large Hadron Collider (LHC), one can search for DM particles ($\chi$) that are pair produced in $pp$ collisions.
These studies are sensitive to low DM masses ($m_{\chi}\le$ 10 GeV), and therefore  provide  
information complementary to direct DM searches, which are most sensitive to larger DM masses ~\cite{lux,cdmsSlite,coupp,picasso}. 
  
If the particles that mediate the interactions between DM and Standard Model (SM) particles
 are too heavy to be produced directly in the expe\-ri\-ment,  their interactions can be described
by  contact operators in the framework of an effective field theory~\cite{Beltran:2010ww,Fox:2011pm,Goodman:2010ku}. 
For each operator considered,  
the reach is expressed in terms of the effective mass scale of the interaction, $M_*$, and 
of the $\chi$--nucleon cross-section, $\sigma_\textrm{\tiny $\chi$-N}$, as a function  of $m_\chi$.

Since DM particles do not interact in the detector, 
the main signature of DM pair production at colliders is large 
missing transverse momentum. 
Initial-state radiation (ISR) of jets, photons, $Z$, or $W$ bosons, was used to tag DM pair production at colliders 
in several searches at the Tevatron~\cite{TevatronMonoJet} 
and the LHC~\cite{ATLASmonojet,CMSmonojet,ATLASmonophoton,CMSmonophoton,ATLASmonoWZ,ATLASmonoZ,ATLASmonolepton, Khachatryan:2014tva, Khachatryan:2014rra}. 

\begin{figure}[tbp]
  \centering \begin{tabular}{c@{\hskip 5mm}c}
  \includegraphics[width=.22\textwidth]{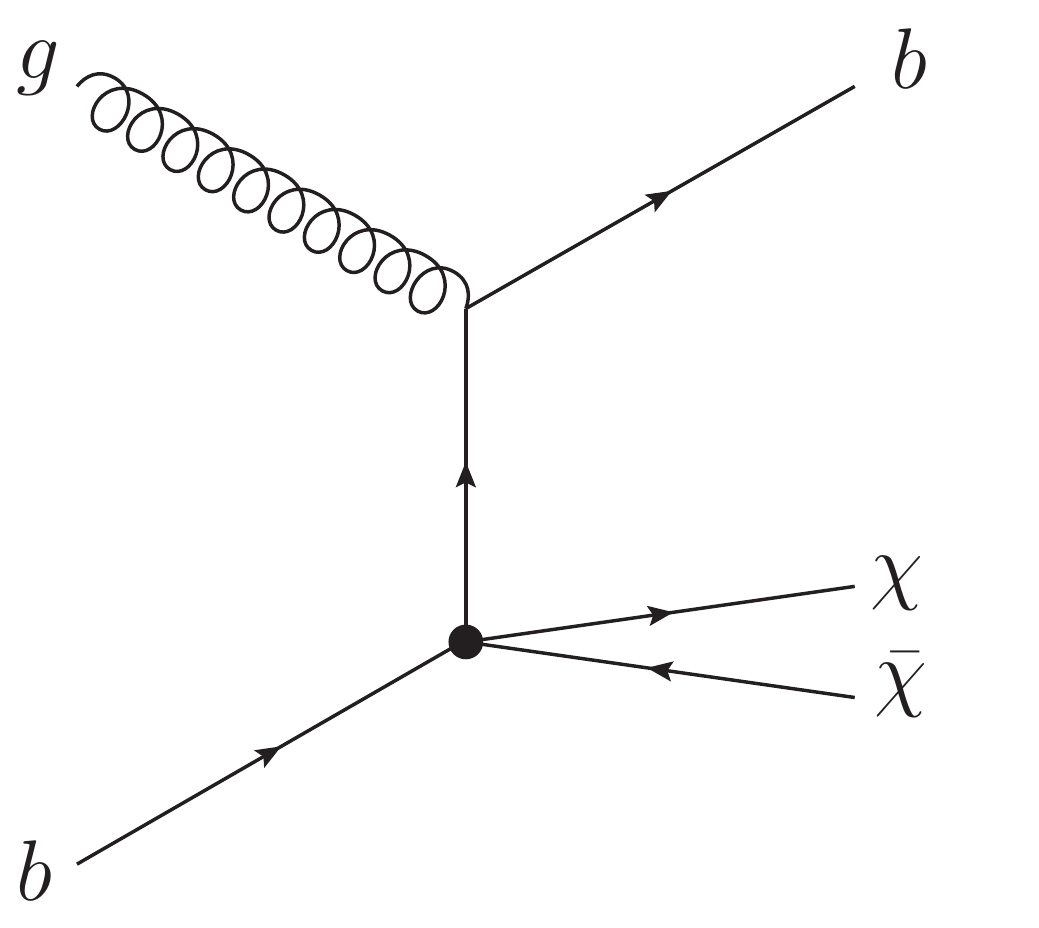}
&
  \includegraphics[width=.22\textwidth]{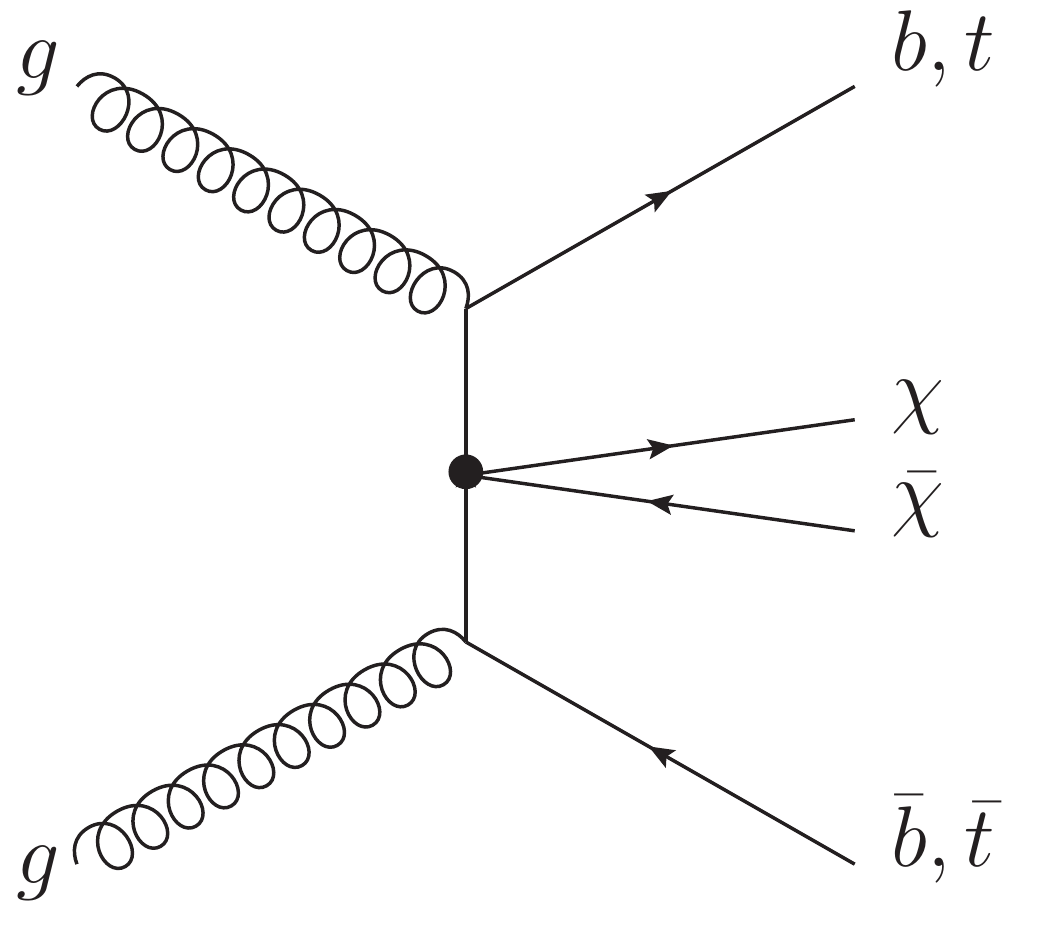}
  \\
  (a) & (b) \\
  \end{tabular}
  \caption{
    Dominant Feynman diagrams for DM production in conjunction with (a) a single $b$-quark and (b) a heavy quark (bottom or top) pair using an effective field theory approach.
  }
  \label{feynman}
\end{figure}

A new search for DM pair production in association with one $b$-quark or a pair of heavy quarks ($b$ or $t$) was proposed in 
Ref.~\cite{Lin:2013sca}.  The dominant Feynman diagrams for these processes are shown in Fig.~\ref{feynman}.
To search for these processes, dedicated selections are defined to reconstruct the various production and decay modes of these heavy-quark final states. 
For final states containing a semileptonic decay of a top quark, the results of the search for a supersymmetric partner of the top quark are used~\cite{stop}.

\begin{figure}[tbp]
  \centering 
  \includegraphics[width=.35\textwidth]{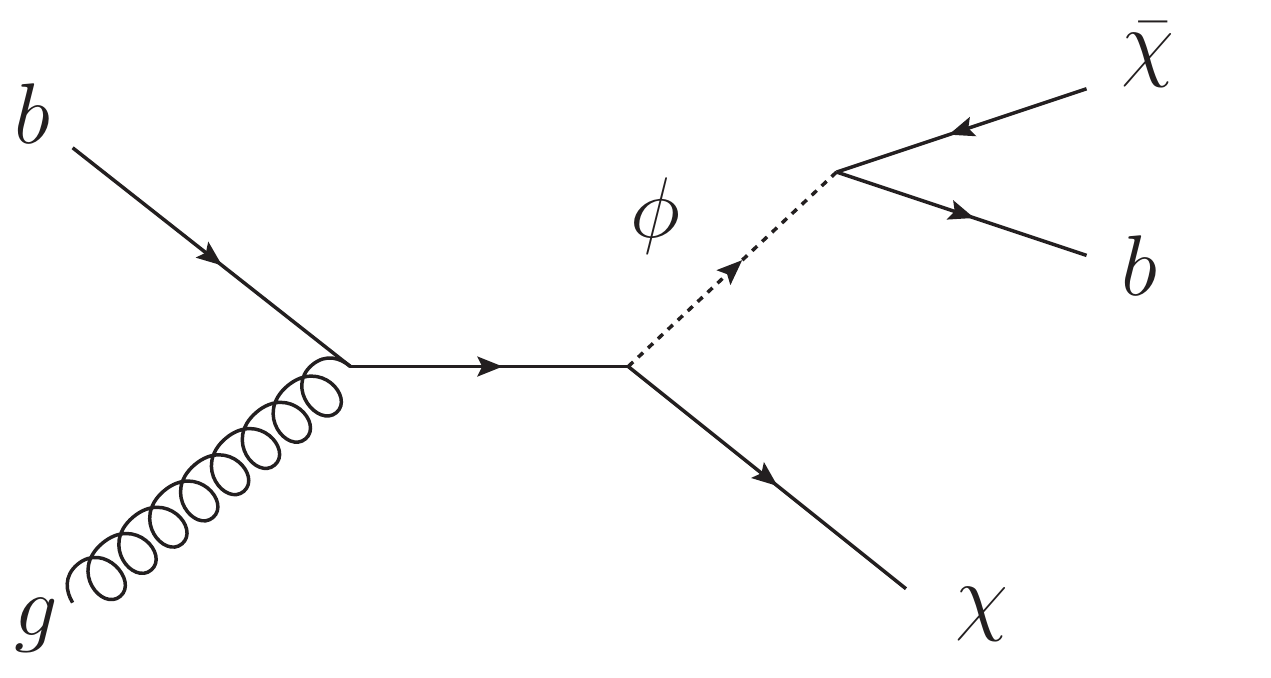}
  \caption{
    Example of DM production in the $b$-FDM model.
  }
  \label{feynman_bfdm}
\end{figure}



The analysis presented in this article is particularly sensitive to  
effective scalar interactions between DM and quarks described by the operator~\cite{Goodman:2010ku} 
\begin{equation}
\phantom{xxxxxxxxxxx}  {\cal O}_\textrm{scalar}= \sum_q \frac{m_q}{M_*^N}\bar{q}q\bar{\chi}\chi , 
\end{equation}

\noindent where $N=3$ for Dirac DM (D1 operator) and $N=2$ for complex scalar DM (C1 operator). 
The quark and DM fields are denoted by $q$ and $\chi$, respectively.
The scalar operators are normalized by $m_q$, which mitigates contributions to flavour-changing processes, strongly constrained by flavour physics observables~\cite{DAmbrosio:2002ex, Buras:2000dm},
through the framework of minimal flavour violation (MFV). The dependence  on the quark mass 
makes final states with bottom and top  quarks the most sensitive to these operators.

This search is also sensitive to tensor couplings between DM and quarks.  
The tensor  operator (D9), which describes a magnetic moment coupling, 
is parameterized as~\cite{Goodman:2010ku}:

\begin{equation}
\phantom{xxxxxxxxx}           {\cal O}_\textrm{tensor} = \sum_q \frac{1}{M_*^{2}}\bar{\chi}\sigma^{\mu\nu}\chi\bar{q}\sigma_{\mu\nu}q.  
\end{equation}
MFV suggests that the D9 operator should have a mass dependence from Yukawa couplings although canonically this is not parametrised as such.

The results are also interpreted in light of  
a bottom-Flavoured Dark Matter model ($b$-FDM)~\cite{Agrawal:2014una}. 
The $b$-FDM model was proposed to explain
the excess of gamma rays from the galactic centre, 
recently observed by the Fermi Gamma-ray Space Telescope, 
and interpreted as a signal for DM annihilation~\cite{Daylan:2014rsa}.  
This analysis of the data recorded by the Fermi-LAT collaboration favours DM with a mass of approximately 35~GeV annihilating into   
$b$-quarks via a coloured mediator. 
In this model, a new 
scalar field, $\phi$, mediates the interactions 
between DM and  quarks as shown in Fig.~\ref{feynman_bfdm}. 
DM is assumed to be a Dirac fermion that couples to right-handed, down-type 
quarks. The lightest DM particle, which constitutes cosmic DM, 
pre\-fe\-ren\-tially couples to $b$-quarks.  
%
%
The collider signature of this model is $b$-quarks produced in association with 
 missing transverse momentum.  
This analysis sets constraints on the mass of the mediator and DM particle in the 
framework of the $b$-FDM model. 

\begin{table*}[htbp]
  \centering
  \caption{Selections for signal regions 1--4. Variables $\pt^{j_i}$ ($\pt^{b_i}$) represent the transverse momentum of the $i$-th jet ($b$-tagged jet). The asymmetric transverse mass $\amt$\cite{Bai:2012gs, Barr:2009jv, Konar:2009qr}, $topness$~\cite{Graesser:2012qy}, $m_{jjj}$~and Razor $R$~\cite{razor} are used to reject the abundant top quark background.}

  \label{tab:selection}
  \begin{tabular}{l|c|c|c|c}\hline\hline
   & SR1 & SR2 & SR3 & SR4\\ \hline
    Trigger                          & \etmiss & \etmiss & $5\ $jets $ ||\ 4 $jets$\left(1b\right)$ & $\etmiss\ ||\ 1$ lepton (no $\tau$)\\ \hline
   Jet multiplicity   $n_j$                   & 1--2 & 3--4 & $\ge$5 & $\ge$4\\ \hline
   $b$-jet multiplicity   $n_b$                   & $>$0   (60\% eff.)   & $>$0 (60\% eff.) & $>$1 (70\% eff.) & $>$0 (70\% eff.)\\ \hline
   Lepton multiplicity   $n_\ell$                         & 0 & 0 &0 & 1 $\ell$ ($\ell=e, \mu$)\\ \hline
   \etmiss                          & $>$300 GeV & $>$300 GeV & $>$200 GeV & $>$270 GeV\\ \hline
   \multirow{2}{*}{Jet kinematics}                   & \multirow{2}{*}{$\pt^{b_{1}}>100$ GeV} & $\pt^{b_{1}}>100$ GeV & \multirow{2}{*}{$\pt^{j}>25$ GeV} & $\pt^{b_{1}}>60$ GeV\\ 
                      &                & $\pt^{j_2}>100$ $\left(60\right)$ GeV         &  & $\pt^{1{\textrm -}4}>80, 70, 50, 25$ GeV\\ \hline
   Three-jet invariant mass &                &                                               &  & $m_{jjj}<360$ GeV\\ \hline
   $\Delta\phi\left(j_{i},\etmiss\right)$ & $>1.0,\ i=1,2$  & $>1.0,\ i=1-4$ & - & $>0.6,\ i=1,2$\\ \hline
   \multirow{3}{*}{Angular selections }     & \multirow{3}{*}{-}& \multirow{3}{*}{-} &\multirow{3}{*}{$\Delta\phi\left(b_{1},\etmiss\right)\ge1.6$} & $\Delta\phi\left(\ell,\etmiss\right)>0.6$\\ 
                      & & & &$\Delta R\left(\ell,j_{1}\right)<2.75$ \\ 

                     & & & & $\Delta R\left(\ell,b\right)<3.0$\\ \hline
   Event shape                       & - &- & Razor $R>$0.75 & $topness>$2\\ \hline
   \amt                              & - &- & - &$>$190 GeV\\ \hline
  $m_{\rm T}^{\ell + E_{\rm T}^{\rm {miss}}}$& - &- & - &$>$130 GeV\\ \hline
   $\met / \sqrt{H_\textrm{T}^{4j}} $& - &- & - &$>$9 $\sqrt{\textrm{GeV}} $\\ \hline\hline


   \end{tabular}
\end{table*}


\section{Detector description and physics objects}

The ATLAS detector~\cite{ATLASdetector}  at the LHC covers the pseudorapidity\footnote{ATLAS uses a right-handed coordinate system with its origin 
at the nominal interaction point (IP) in the centre of the detector, 
and the $z$-axis along the beam line. The $x$-axis points from the IP to
the centre of the LHC ring, and the $y$-axis points upwards. Cylindrical coordinates
($r,\ \phi$) are used in the transverse plane, $\phi $ being the azi\-mu\-thal angle around the beam line.
The pseudorapidity $\eta$ is defined
in terms of the polar angle $\theta $ as $\eta=-\ln\tan(\theta/2)$.
Observables labeled ``transverse'' are projected into the $x$--$y$ plane.} 
range of $| \eta | <$ 4.9 and is hermetic in azi\-muth $\phi$. 
It consists of an inner tracking detector surrounded by a superconducting solenoid, electromagnetic and hadronic calorimeters, 
and an external muon spectrometer incorporating large superconducting toroidal magnets. 
A three-level trigger system is used to select  events for 
subsequent offline analysis. 
The data set used in this analysis consists of \Lumi~of $pp$ collision data  
recorded at a centre-of-mass energy of $\sqrt{s}$ = 8~TeV with stable beam 
conditions~\cite{LumiRef} during the 2012 LHC run. 
All subsystems listed above were required to be operational.

This analysis requires the reconstruction of  muons, electrons, jets, and missing transverse momentum. 
%
%
Muon candidates are identified from tracks 
that are well recon\-structed inside both the inner detector and the muon 
spectrometer~\cite{combinedmuons}. 
%
To reject cosmic-ray muons, muon candidates are required to be consistent 
with production at 
the primary vertex,  defined as the vertex with the highest $\Sigma (p_\mathrm{T}^{\mathrm{track}} )^2$, where $p_\mathrm{T}^{\mathrm{track}}$ refers to the transverse momentum of each track.

%
Electrons are identified as 
tracks that are matched to a well-reconstructed  cluster in the electromagnetic calorimeter.  
Electron candidates must satisfy the {\it tight} electron shower shape and track selection criteria of Ref.~\cite{mediumelectrons}. 
Both electrons and muons are required to have transverse momenta $\pt\ > 20$ 
GeV and $|\eta |<2.5$.  
Potential ambiguities between overlapping candidate objects are resolved based on their angular separation.
If an electron candidate and a jet overlap within $\Delta R< 0.2$, then the object is considered to be an electron and the jet is discarded. If an electron candidate and any jet overlap within $0.2 < \Delta R < 0.4$, or if an electron candidate and a b-tagged jet overlap within $\Delta R< 0.2$ of each other, then the electron is discarded and the jet is retained.
  
Photon candidates must satisfy the {\it tight} quality criteria 
and $|\eta| < 2.37$ \cite{Aad:2014nim}.

Jet candidates are reconstructed  using the anti-$k_t$ clustering algorithm~\cite{paper:antikt} 
with a radius parameter of 0.4. 
The inputs to this algorithm are three-dimensional topological clusters~\cite{topocluster}.
The four-momentum of the jet is defined as the vector sum of the four-momenta of the topological clusters, 
assuming that each cluster originates from a particle defined to be massless and to come from the interaction point.

To calibrate the reconstructed energy, jets are corrected for the effects of calorimeter response and inhomogeneities using energy- and $\eta$-dependent calibration factors based on simulation and validated with extensive test-beam and collision-data studies~\cite{Aad:2011he}. In the simulation, this procedure calibrates the jet energies to those of the corresponding jets constructed from stable simulated particles. In-situ measurements are used to further correct the data to match the energy scale in simulated events. Effects due to additional $pp$ interactions in the same and preceding bunch crossings (pile-up effects) are corrected~\cite{Aad:2014bia}.
Only jets with $\pt >$20 (25) GeV and $|\eta |<$4.5 (2.5) are considered in this analysis for final states involving $b$ ($t$) quarks.

Jets containing particles from the hadronisation of a $b$-quark \mbox{($b$-jets)} are tagged using a multivariate algorithm~\cite{2014:004, ATLAS-CONF-2014-046}.
The $b$-tagging algorithm combines 
the measurement of several quantities distinguishing heavy quarks from light quarks based on their longer lifetime and heavi\-er mass. 
These quantities include the 
distance of closest approach of tracks in the jet to the primary event vertex,
the number and position of secondary vertices formed by tracks within the jet, as well as the invariant mass 
associated with such vertices. 
The algorithm is trained on Monte Carlo (MC) simulations and its performance is calibrated using data. 
To optimize the sensitivity of this analysis, a requirement on the output of the $b$-tagging algorithm which provides a 60\% (70\%) $b$-jet efficiency operating point is used in signal regions
(SR) 1 and 2 (3 and 4) defined below.
The corresponding misidentification proba\-bi\-li\-ty is 15\% (20\%) for $c$-jets, 
and less than 1\% for 
light-quark jets.  
The aforementioned \mbox{$b$-tagging} efficiencies and misidentification probabilities 
were derived in a simulated \ttbar\ sample with jet transverse momenta of $\pt>20$~GeV and $|\eta|<2.5$.

The missing transverse momentum, with magnitude \etmiss, is defined as the negative vector sum of the transverse 
momenta of jets, muons, electrons, photons, and topological clusters not assigned to any reconstructed objects~\cite{etmissref}. 

\section{Event selection}

\begin{table*}[tbp]
  \centering
  \caption{
Expected  background and signal yields for $m_\chi$=10~GeV compared with observed yields in data for the various signal regions. For the $b$-FDM model, $m_\phi$ is 600~GeV. The row labeled ``Total expected background" shows the sum of all background components. 
The quoted uncertainties include all statistical and systematic effects added in quadrature.
The effective mass scale, $M_*$, is set to be 100/40/600 GeV for the D1/C1/D9 operators, approximately corresponding to the expected limit.
 The probabilities of the background-only hypothesis, $p$-values, are also given. The last two lines show the observed and expected 95\% CL upper limits on the number of beyond-the-SM events.
 }
  \label{tab1}
  \newcommand{\twrw}[1]{\multirow{2}{*}{#1}}
  \begin{tabular}{l|c|c|c|c}\hline \hline
  {\bf Background source}                & {\bf SR1}              & {\bf SR2}          &   \bf{SR3}          &      \bf{SR4}       \\\hline
  $Z$($\nu\overline{\nu}$)+jets          & 190 $\pm$ 26           & \ph90 $\pm$ 25\ph  & \multirow{2}{*}{$1^{+6}_{-1}$}   & --              \\ \cline{1-3}\cline{5-5}
  $W$($\ell\nu$)+jets                    & 133 $\pm$ 23           & \ph75 $\pm$ 13     &                     & 1.3 $\pm$ 0.3                        \\ \hline
   \ttbar \                              & \multirow{2}{*}{39 $\pm$ 5}                 & \multirow{2}{*}{\ph71 $\pm$ 9} & 87 $\pm$ 11 &2.9 $\pm$ 0.6\\ \cline{4-5}
  Single top                                                                & &        & 8 $\pm$ 3           & 0.7 $\pm$ 0.3  \\\hline
  \ttbar +$Z$/$W$                        & ~--                    & \ph--              & --                  & 1.4 $\pm$ 0.4  \\\hline
  Diboson                                & 22 $\pm$ 4             & \ph8 $\pm$ 1       & --                  & 0.8 $\pm$ 0.4  \\  \hline
  {\bf Total expected background}              & 385 $\pm$ 35     & 245 $\pm$ 30       & 96 $\pm$ 13         & 7 $\pm$ 1  \\   \hline  
  {\bf Data}                             & 440 & 264 & 107 & 10\\   \hline   \hline
  {\bf Expected signal} -- D1            & 10 $\pm$ 2       &   \ph49 $\pm$ 8\ph       &   28 $\pm$ 2\ph     &  35 $\pm$ 5\phantom{x} \\ \hline
  {\bf Expected signal} -- C1            & 17 $\pm$ 2       &   \ph61 $\pm$ 9\ph       &   45 $\pm$ 4\ph     &  51 $\pm$ 12  \\ \hline
  {\bf Expected signal} -- D9            & 147 $\pm$ 25     &   \ph69 $\pm$ 12         &  \ph2 $\pm$ 1\ph    &   2 $\pm$ 1\\ \hline
  {\bf Expected signal} -- $b$-FDM       & 192 $\pm$ 24     &   \ph61 $\pm$ 8\ph       &  1.0  $\pm$ 0.2     &  --  \\ \hline \hline
  {\bf $p$-Value}                        & 0.09 & 0.29 & 0.24 & 0.18\\   \hline   
  Allowed non SM events - Obs.           & 124 & 79 & 41& 10\\   \hline   
  Allowed non SM events - Exp.           & 81 & 67 & 33& 7\\   \hline \hline
   \end{tabular}
\end{table*}


Candidate signal events containing at least one high-\pt\ jet and large \etmiss\ are assigned to one of four orthogonal signal regions. 
The first two signal regions focus on events with DM produced in conjunction with one (SR1) or two (SR2) $b$-quarks 
in the final state. SR3 and SR4 target events in which DM is produced in conjunction with a $t\overline{t}$ pair, 
where either both top quarks decay hadronically (SR3) or 
one top quark decays hadronically and the other  semileptonically (SR4). 
SR4 was developed for a top squark search by the ATLAS Collaboration and  
coincides with  the ``tNbC\_mix'' signal region described in Ref.~\cite{stop}.  
 The four signal regions 
provide the complementary information needed in case of observation of a signal.

Events assigned to SR1 and SR2  are  required to pass 
a calorimeter-based \etmiss\ trigger with a threshold of 80~GeV.  
To enrich the sample in $pp \to \chi\bar{\chi}+b(\bar{b})$, 
events are required to have a low jet multiplicity ($n_{\textrm{jets}}<$5),  \etmiss $>$ 300 GeV, and
the most energetic $b$-tagged jet must 
have a \pt $>$ 100~GeV.
The azi\-mu\-thal separation between the directions  of the jets and the missing transverse momentum is required to be more than 1.0 radian.  
Events with at least one identified muon or electron are discarded to reject leptonic decays of $W$ and $Z$ bosons. 
Events satisfying these selection criteria are assigned to SR1 provided that the jet multiplicity does not exceed two. 
Events are assigned to SR2 when at least three jets are reconstructed in the event 
and the second most energetic jet  has $\pt > 100$~GeV. If there is a second $b$-tagged jet it has to satisfy $\pt > 60$ GeV.

Events assigned to SR3 are required to pass 
triggers specifically designed to select hadronic decays of top quark pairs. 
Such triggers require either five jets with $p_\mathrm{T}\ge$ 55 GeV  
each or four jets with $p_\mathrm{T}\ge$ 45~GeV, 
of which one is tagged as a $b$-jet.
%
To select $pp \to \chi\bar{\chi}+t\bar{t} $ events, at least five reconstructed jets are required, of which at least two are $b$-tagged, and $\etmiss >  200$ GeV. Furthermore, the azimuthal separation between the most energetic $b$-jet and the missing transverse momentum is required to be at least 1.6 radians.  
To reduce $W/Z$ leptonic decays and leptonic top quark decays, 
events with at least one identified muon or electron are discarded. 
To maximize the rejection of the abundant \ttbar\ background, the Razor variable $R$~\cite{razor} is used. 
This variable utilizes both transverse and longitudinal information about the event 
to fully exploit the kinematics of the decay.
%
%
To separate signal and background, $R >$ 0.75 is required.

\begin{figure*}[!htbp]
  \centering \begin{tabular}{c@{\hskip 5mm}c}
   \includegraphics[width=.45\textwidth]{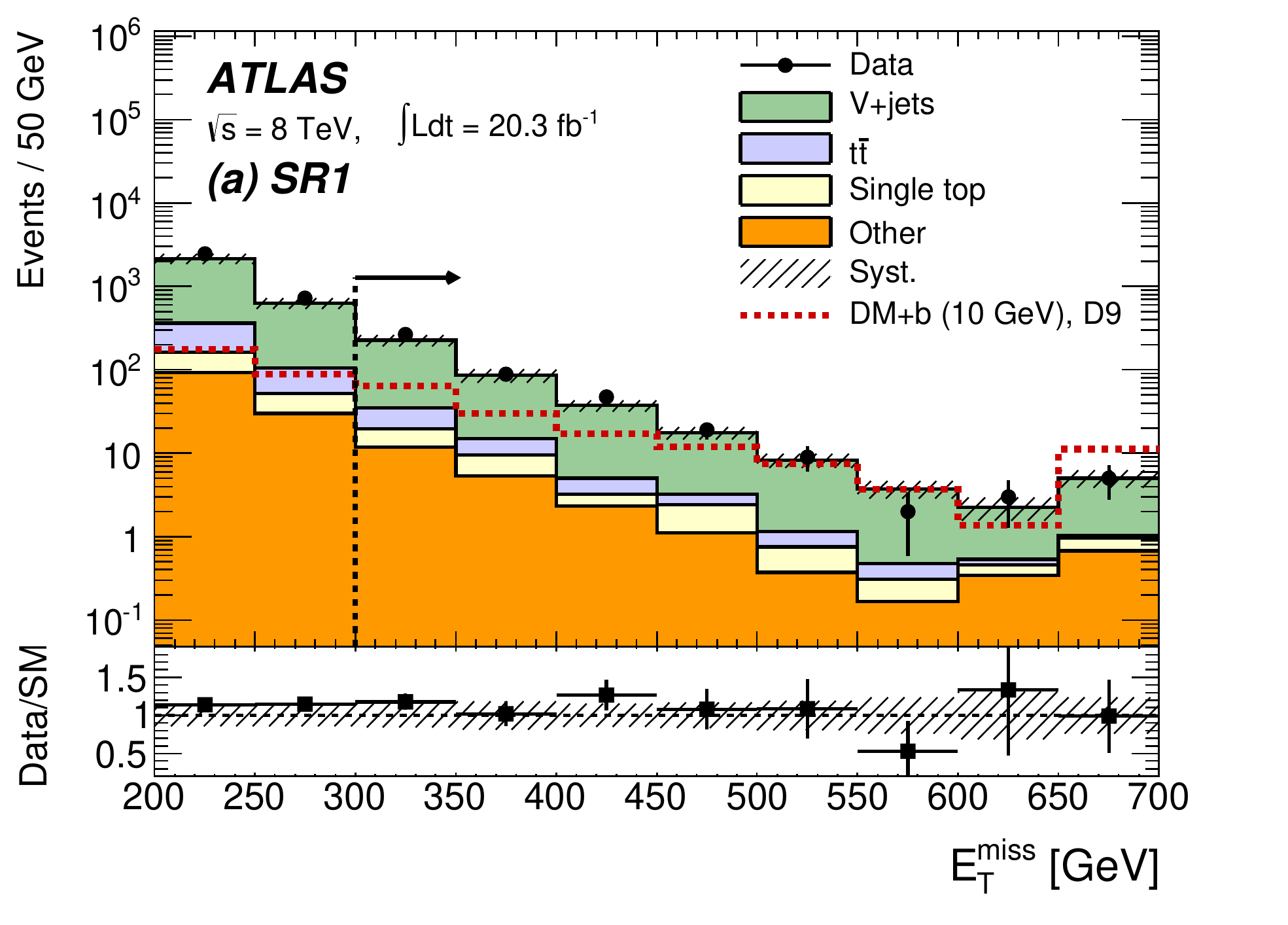}
   \includegraphics[width=.45\textwidth]{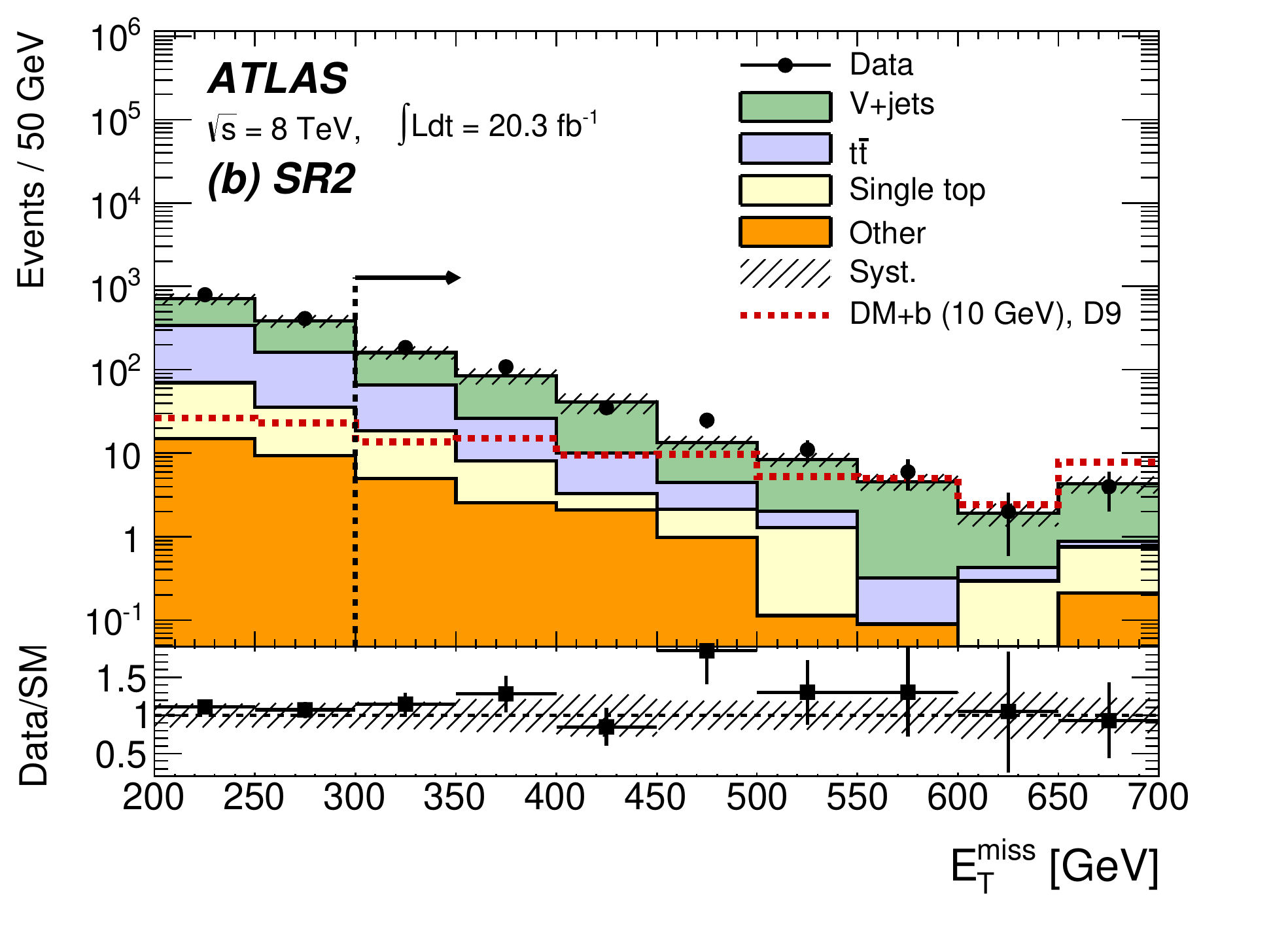}\\
   \includegraphics[width=.45\textwidth]{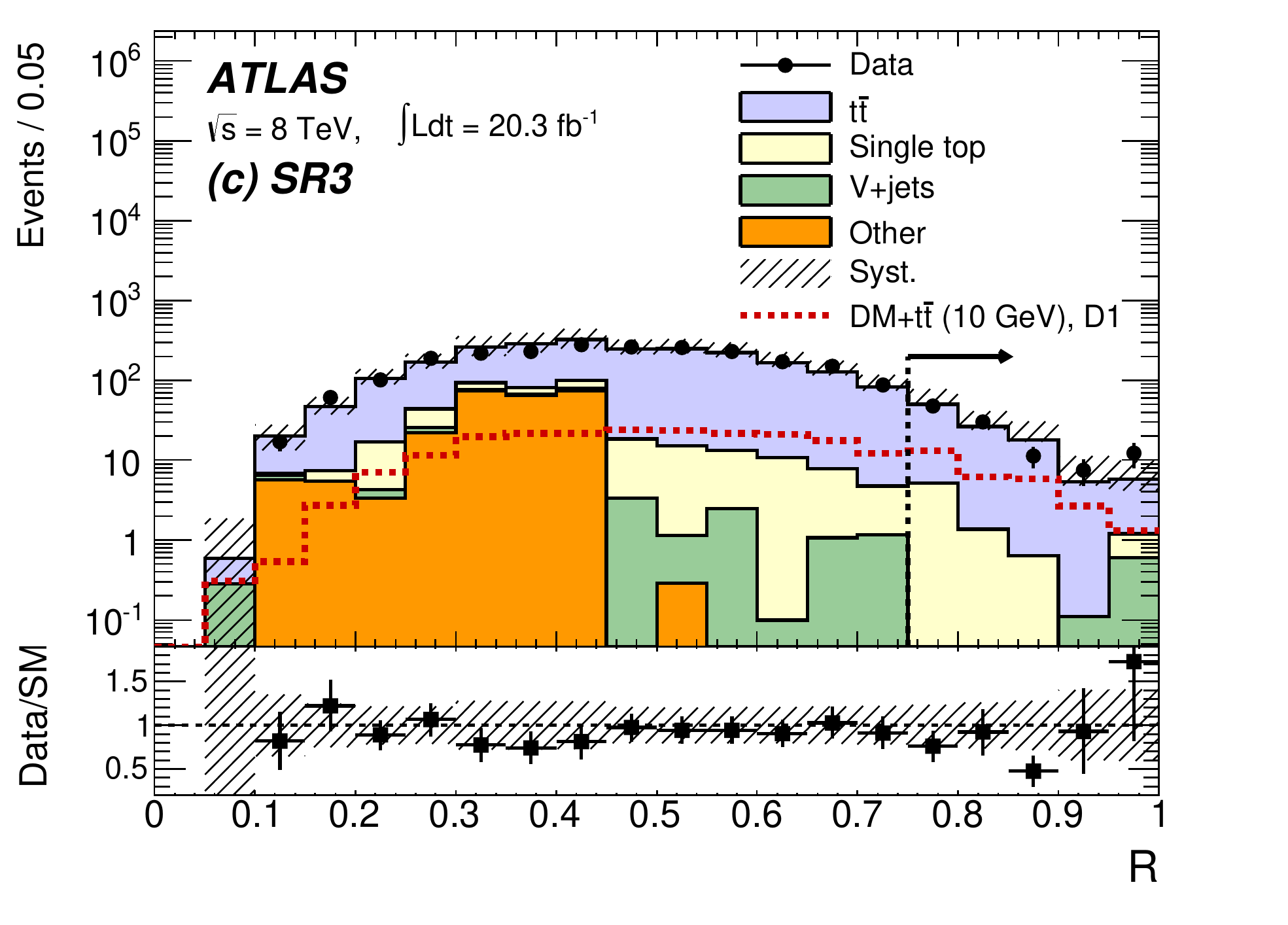} 
   \includegraphics[width=.45\textwidth]{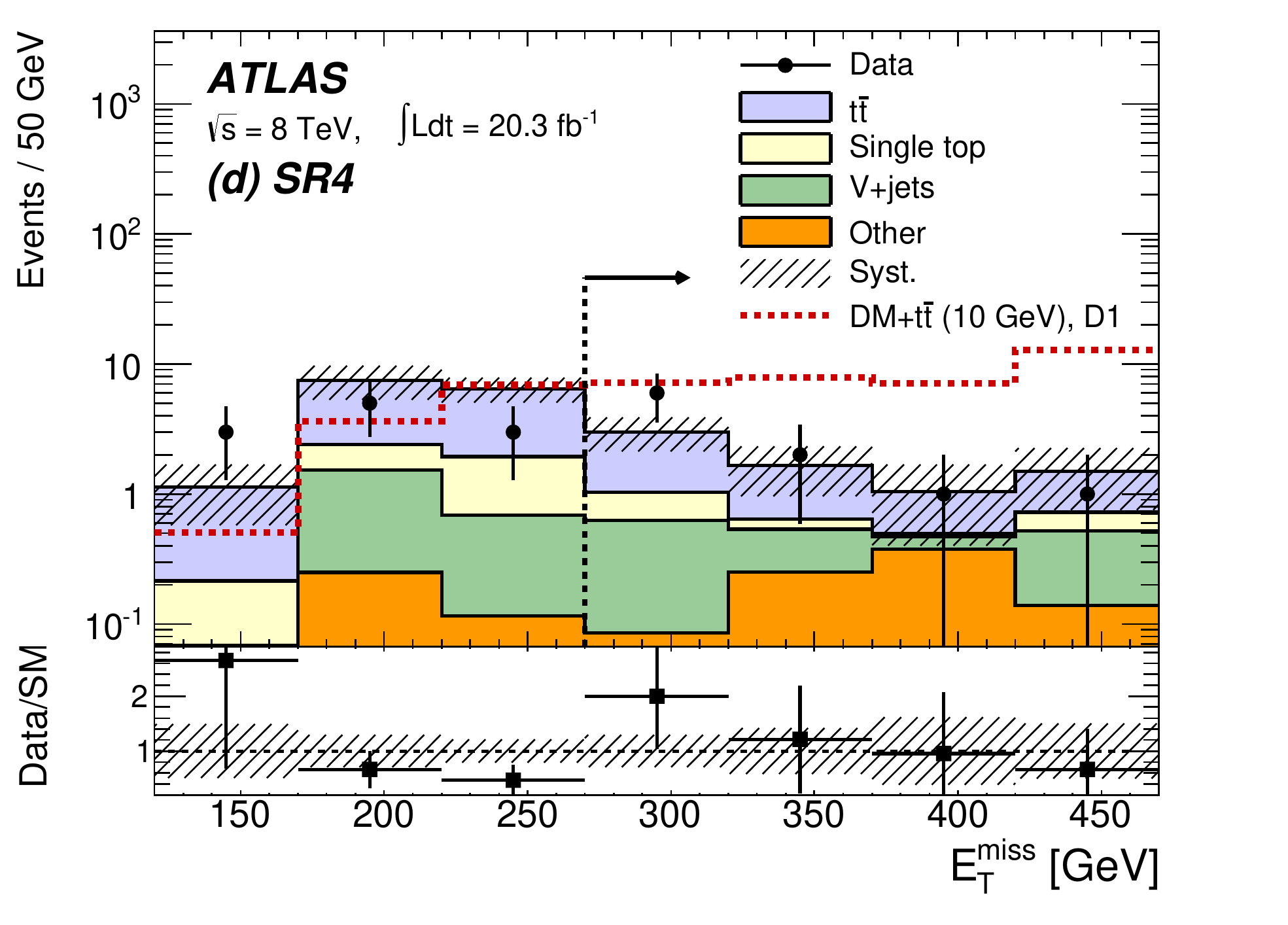} \\
   \end{tabular} 
  \caption{Comparison between data and expected SM background.  
    (a), (b): \etmiss\ variable for SR1 and SR2 
    and for an example signal with the operator D9. 
    (c): $R$ variable for SR3 excluding the selection on $R$ and for an example signal with the operator D1. 
    (d): \etmiss\ variable for SR4 excluding the selection on \etmiss and for an example signal with the operator D1. Other backgrounds are composed of diboson and multijet production.  The expected signal for $\chi \bar{\chi} + b(\bar{b})$ (SR1, 2) and for
    $\chi \bar{\chi} + t\bar{t}$ (SR3, 4) production for $m_\chi=$~10~GeV is given by the red line assuming $M_*$ =  100/40/600 GeV for the D1/C1/D9 operators, respectively.  
    The final selection requirements are indicated by an arrow.
    The error bars represent the statistical uncertainty. The dashed area shows the systematic uncertainty
    on the background estimation. Events with values exceeding the range presented are included in the highest bin.
 }
  \label{DataMC}
\end{figure*}

To enrich the sample in $pp \to \chi\bar{\chi}+t\bar{t}$ with one 
semileptonic decay of the $t$ quark, 
events assigned to SR4 
use single-lepton or \etmiss\ triggers,  
and require exactly one isolated lepton (electron or muon) with $\pt > 25$~GeV, 
at least four high-\pt\ jets, where one jet is $b$-tagged with  $\pt > 60$~GeV. 
Events with $\etmiss>270$ GeV are selected when 
 the transverse mass\footnote{Since the longitudinal component of the momentum of the neutrinos is not measured, the measured properties of the $W$ boson candidates are limited to their transverse momentum and transverse mass, defined as $m_\mathrm{T}=\sqrt{ (\etmiss + p_\mathrm{T}^\ell)^2 - (E^{\mathrm{miss}}_x\! + p_x^\ell)^2  - (E^{\mathrm{miss}}_y\! + p_y^\ell)^2}$ where \etmiss\ is the magnitude of the missing transverse momentum vector, $p_\mathrm{T}^\ell$ is the transverse momentum of the lepton and $p_x^\ell$ and $p_y^\ell$ ($E^{\mathrm{miss}}_x\!$ and $E^{\mathrm{miss}}_y\!$) are the magnitude of the $x$ and $y$ components of the lepton momentum (missing transverse momentum) respectively.
}
 formed by the lepton and \met,
 $m_\mathrm{T}(\ell,\met)$, exceeds  130~GeV and 
 $\met / \sqrt{H_\textrm{T}^{4j}} > 9~\sqrt{\textrm{GeV}}$, 
with $H_\textrm{T}^{4j} = \sum_{i=1}^4 \pt(\textrm{jet}_i)$ and 
where the jets are ordered by decreasing \pt.
The azimuthal angle between the missing transverse momentum and the two most energetic 
jets is required to be greater than 0.6 radians. 
%

Special variables, such as the asymmetric transverse mass $\amt$ \cite{Bai:2012gs, Barr:2009jv, Konar:2009qr} 
and the $topness$ variable~\cite{Graesser:2012qy}, are used to reject the dileptonic \ttbar\ component of the background. Details can be found in Ref.~\cite{stop}.
The diboson background is suppressed by a requirement on the three-jet invariant mass ($m_{jjj}<360$ GeV)~\cite{stop}.
A $\tau$ veto rejects \ttbar\ events with hadronically decaying $\tau$ leptons in the final state. Additional selection criteria~\cite{stop} on the angles 
between the lepton and the various jets are imposed to further 
reduce the \ttbar\ background. 
Table~\ref{tab:selection} provides an overview of the selections applied in all four signal regions.

The product of the detector acceptance ${\it A}$ and the reconstruction efficiency $\epsilon$
for the selections described above varies between  0.1\% and 8\% depending on the signal region, 
operator, and  specific channel 
considered. SR1 and SR2 have the highest efficiencies (${\it A} \times \epsilon >2\%$) for the D9 operator, 
while SR3 and SR4 are most efficient for the D1 and C1 operators (${\it A} \times \epsilon >1\%$).

The dominant background for SR1 and SR2 is 
due to $Z\to \nu\overline{\nu}$ events 
produced in conjunction with one or more jets.  
This irreducible background is estimated from data using two control regions (CRs).   
The first CR exploits $Z$+jets events  with $Z\to\mu^+\mu^-$, while 
the second uses 
$\gamma$+jets events 
for which the production at high transverse momentum ($\pt^{\gamma}>M_Z$) mimics 
that of $Z$+jets~\cite{gammajets}. 
The $\gamma$+jets control region  substantially increases the number of events at large missing transverse momentum. 
The transverse momentum of the dimuon pair or photon is added vectorially to the \etmiss\ of the event to 
simulate the $Z\to\nu \bar{\nu}$ background. 
Corrections to compensate  for the differences in efficiency and acceptance between 
the $Z(\nu\overline{\nu})$+jets and $Z(\mu^+\mu^-)$+jets or $\gamma$+jets 
are derived from data using control regions without $b$-tagged jets before applying any requirements on the missing transverse momentum.
Remaining kinematic selections correspond to the ones described in Tab.~\ref{tab:selection}.
 A muon control region is chosen because the energy loss of muons in the detector is comparatively small. 
The systematic uncertainties introduced by this data-driven procedure
on the $Z(\nu\bar{\nu})$+jets background are approximately \sysZnunu\%, mainly from the flavour composition of background processes, kinematic differences between the control and signal regions and relative normalizations of backgrounds.

Production of $W/Z$+jets with subsequent leptonic decays of $W$ and to a much smaller degree $Z$
is also a substantial source of background for SR1 and SR2 when 
the resulting charged leptons fail to be identified or if the $W$ or $Z$ bosons decay to $\tau$ leptons.
These contributions are estimated from 
$Z(\ell^+ \ell^-)$+jets and  $W(\ell \nu)$+jets MC samples 
ge\-ne\-ra\-ted using  ALPGEN2.3~\cite{ALPGEN}  with the CTEQ6L1~\cite{Pumplin:2002vw} parton distribution function (PDF) set. 
The procedure used for the normalization of this sample is described in reference~\cite{Aad:2014zka}.
These samples are ge\-ne\-ra\-ted with up to five light partons ($u,~d,~s$) and one $c$ quark or two heavy quarks ($c$, $b$) per event. $W+b$ production is highly suppressed and therefore negligible. A control region enriched in  $W(\ell \nu)$+jets events is selected by adding a lepton requirement to the selection and
is used to validate the estimate of this background. The purity of $W(\ell \nu)$+jets in the control region for SR1 (SR2) is 67\% (47\%). 
After full selection the contribution of $b$($c$)-quarks to the dominant $W(\ell \nu)$+jets background is approximately 39\% (38\%) for SR1 and 52\% (37\%) for SR2. The systematic uncertainty on this background is approximately 20\%.
Finally, the small contribution from \ttbar\ is estimated using MC samples and validated in data control regions before 
applying signal selection requirements. The \ttbar\ process is selected with very high purity by requiring events with one lepton and large jet multiplicities.

\begin{figure*}[htbp]
      \includegraphics[scale=0.29]{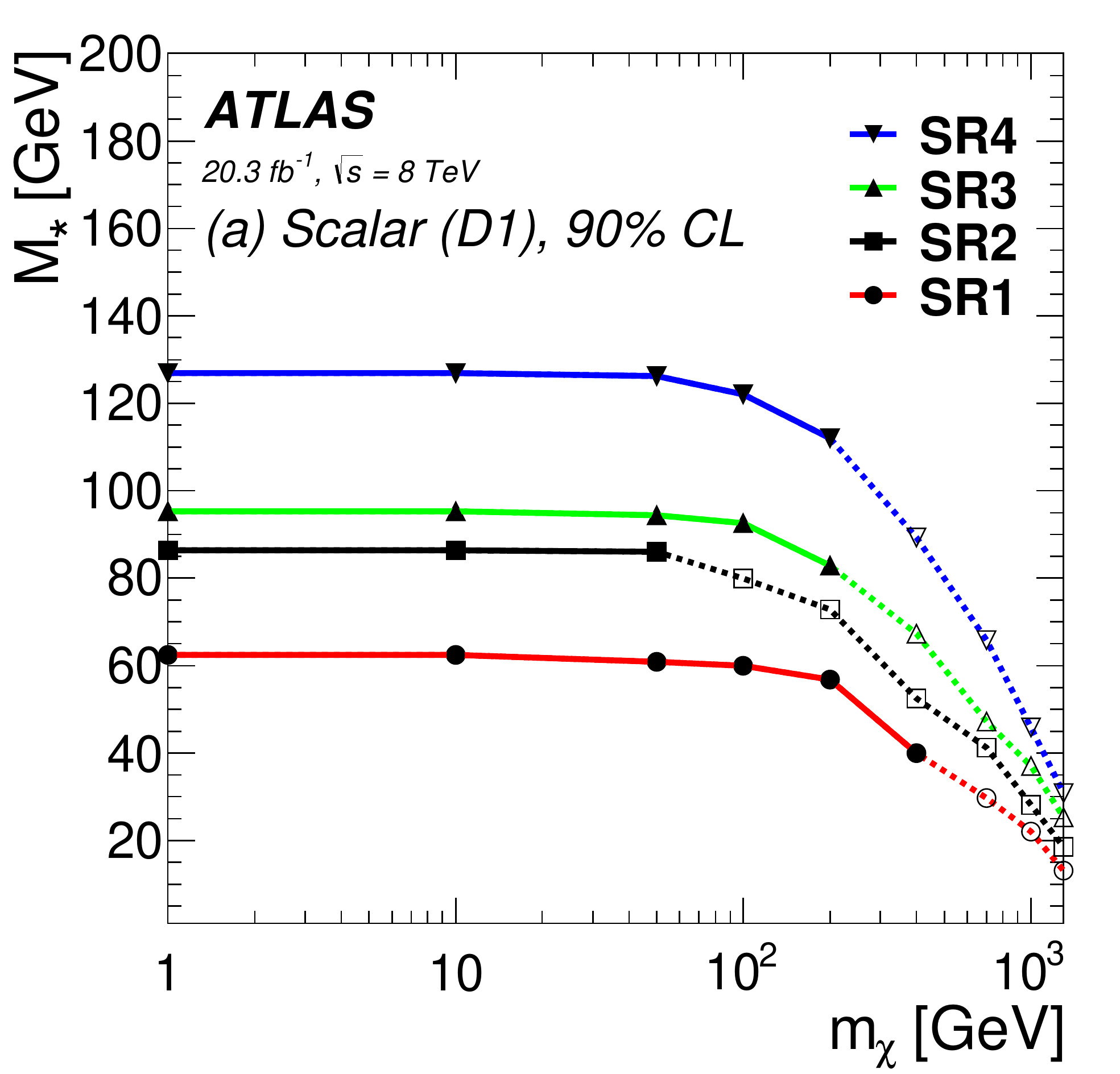}
      \includegraphics[scale=0.29]{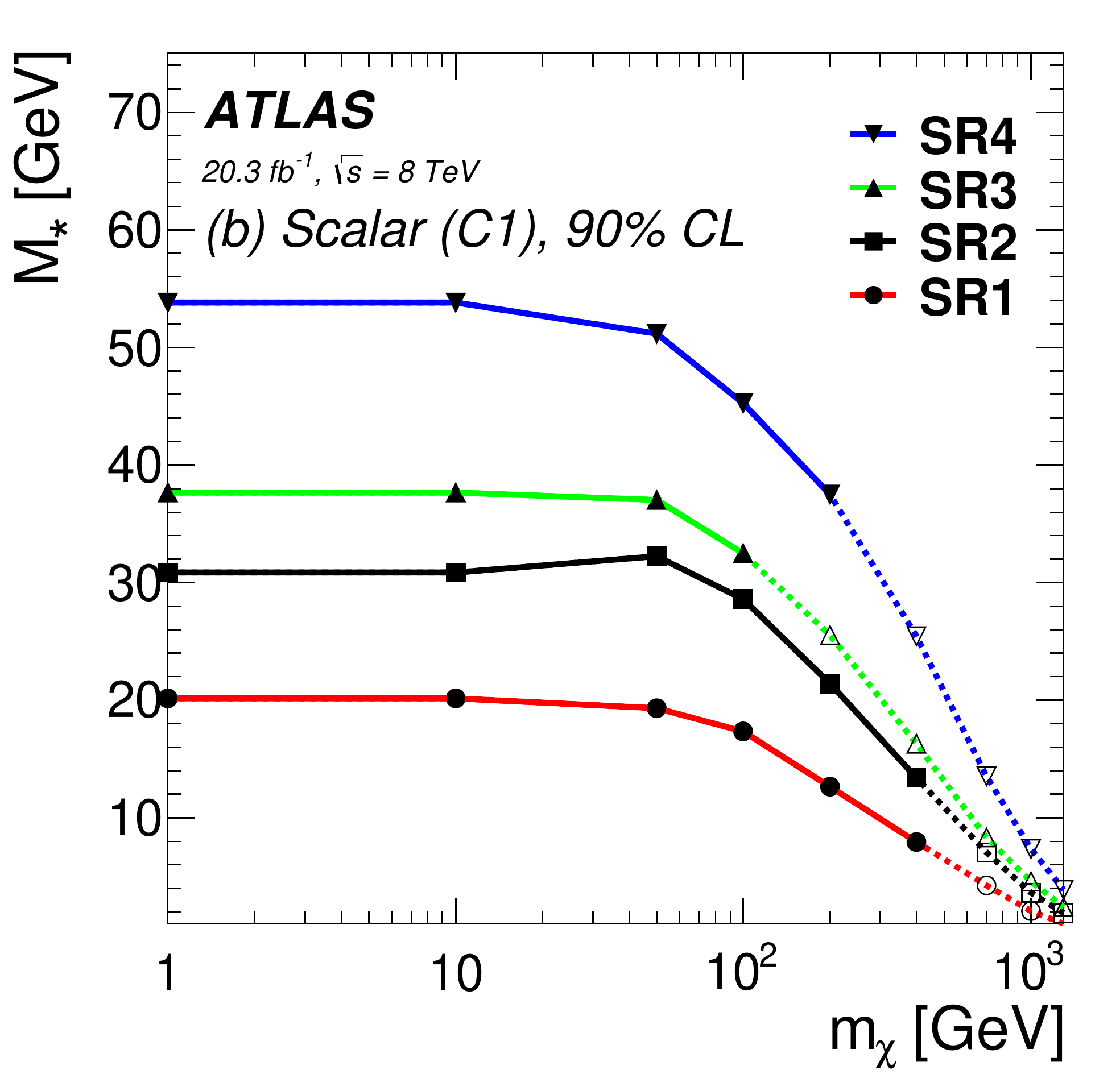}
      \includegraphics[scale=0.29]{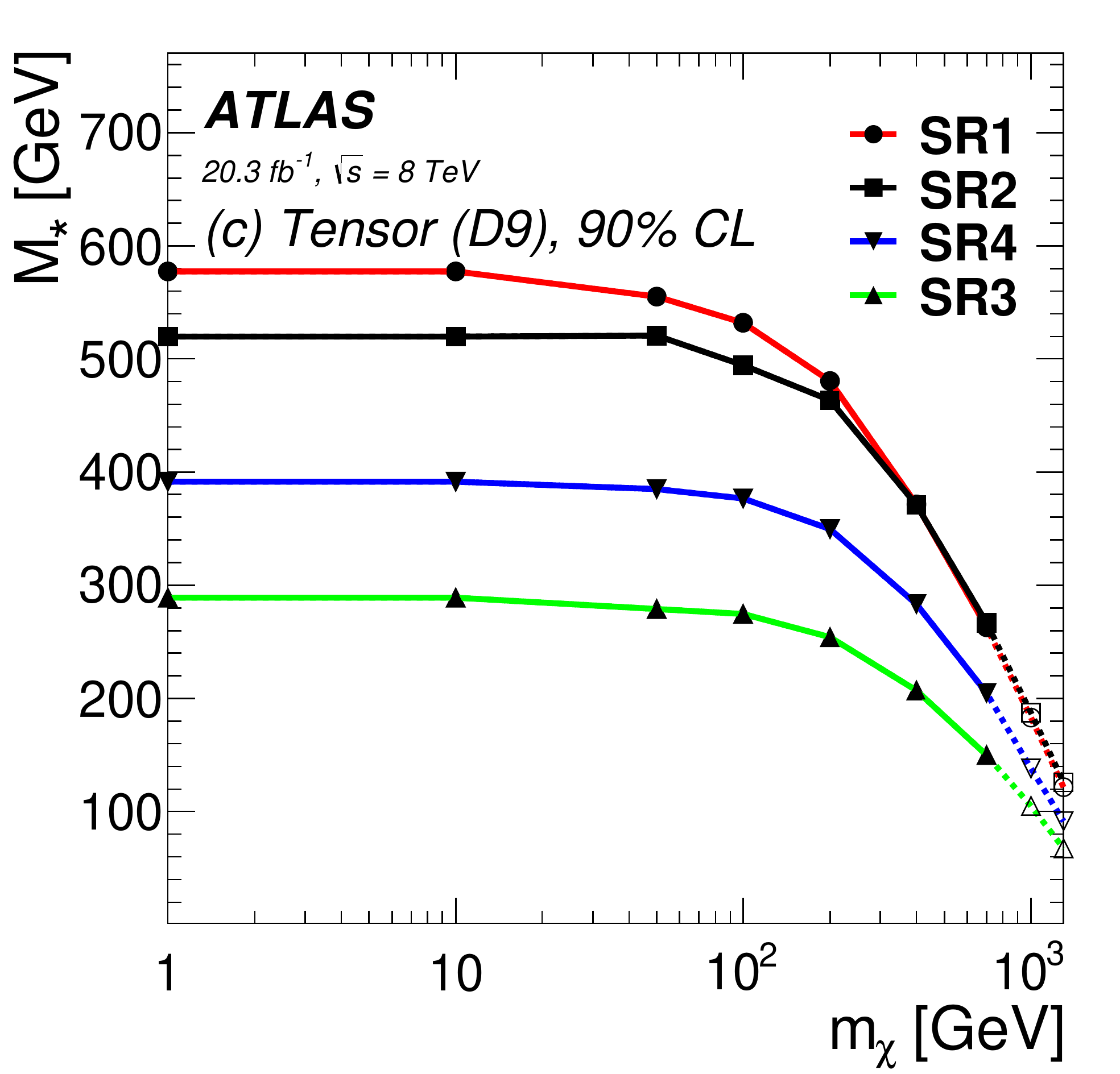}
    \caption{Lower limits on $M_*$ at 90\% CL for the  SR1 (red), SR2 (black), SR3 (green), and SR4 (blue) 
  as a function of $m_{\chi}$ for the operators (a) D1, (b) C1, and (c) D9. Solid lines and markers indicate the validity range of the effective field theory assuming couplings $g_q g_\chi<4\pi$, the dashed lines and hollow makers represent the full collider constraints.
}
  \label{figMstar_90}
\end{figure*}

The dominant source of background for SR3 and SR4 is \ttbar\ events. 
In SR3, this contribution is estimated from data using a control region not overlapping with SR4 and largely dominated by \ttbar\ events with one of the two top quarks 
decaying semileptonically. The five-jets requirement is relaxed to three jets. Additionally, the event is required to contain exactly one lepton with $p_{\rm T}^{e(\mu)}>30~(25)$~GeV and must fullfil $\etmiss+m_\mathrm{T}>25~(30)$~GeV for the electron (muon) channel. The potential signal contribution to this selection is less than 0.1\%. 
The uncertainties are small because the SR3 data control region uses a kinematic region similar to the signal region with the lepton veto and jet multiplicity being the main difference. 
These effects were studied and considered as systematic uncertainties. 
Dominant uncertainties are related to jets and the top quark momentum distribution. 
Corrections to compensate  for the differences in efficiency and acceptance between 
hadronic and semi\-leptonic top decays are derived from MC samples generated using  the POWHEG BOX generator~\cite{powheg} interfaced with  JIMMY4.31~\cite{JIMMY} with the next-to-leading-order (NLO) PDF set CT10~\cite{Lai:2010vv}. The systematic uncertainty on the \ttbar\ background in SR3 of approximately $7\%$ is derived by studying corrections for the top quark momentum distribution, and shower modelling by interfacing the same generator with PYTHIA6 \cite{PYTHIA6,PYTHIA6tune}.

In SR4, the $t\bar{t}$ background is estimated from data using a control region obtained 
by requiring $60~\textrm{GeV}< m_\textrm{T}< 90$ GeV and loosening the selection criteria on \mbox{$\etmiss $},  \amt, 
and \mbox{$\met / \sqrt{H_\textrm{T}^{4j}}$}. A similar selection, but applying an inverted $b$-tagging requirement, is used to estimate the $W$($\ell\nu$)+jets background.
The uncertainty on the \ttbar\ background is estimated to be approximately 20\%~\cite{stop}, which is larger than the uncertainty in SR3 due to the 
limited statistics. 
These uncertainties are evaluated by varying the renormalisation and 
factorisation scale of the simulations, comparing alternative PDF sets, and studying the effects of different shower generators and 
of ISR and final-state radiation. 

Additional sources of background, which include
single-top, $t\bar{t}+Z/W$, and diboson production, are estimated in all signal regions using simulations and  
NLO cross sections~\cite{Campbell:2012dh,Garzelli:2012bn}. 
The single-top (s-channel) and $Wt$ background is generated using the POWHEG generator.
The single-top t-channel is generated with ACERMC3.8~\cite{ACER} interfaced with PY\-THIA6. Associated production of \ttbar\ and a vector boson ($W$, $Z$) are generated with MADGRAPH5~\cite{Alwall:2011uj} with up to two additional partons interfaced with PY\-THIA6. The cross-sections for \ttbar\  production in association with a $W$ ($Z$) boson are determined using the MSTW2008 NLO (CTEQ6.6M) PDF sets.
The diboson samples are generated using HERWIG6.520~\cite{HERWIG,Marchesini:1991ch} and JIMMY4.31 with the CTEQ6L1 PDF set.
The multijet background is estimated using data-driven methods \cite{Aad:2012ms} and is found to be negligible in all signal regions after full selection.

Object reconstruction efficiencies in simulated events are corrected to reproduce the performance measured in data.
The systematic uncertainty of the background estimates derived from simulation 
combines  the  uncertainties  on the 
efficiency of the $b$-tagging algorithm, the uncertainties  on the determination of the 
energy scale and resolution of the jet energy and \etmiss , 
the theore\-ti\-cal uncertainty on the various cross-sections, 
changes in the shapes of distributions used to
extrapolate event counts from control regions to the signal region, data driven corrections and the PDF uncertainties. 
Overall, the systematic uncertainty on the background estimated from simulation
is calculated to be between 12\% and 18\%, depending on the signal region. 

The simulation of the signal samples of  $pp\to\chi\overline{\chi}+ b(\overline{b})$,
$pp\to\chi\overline{\chi}+ t\overline{t}$,  and $b$-FDM
employs the MADGRAPH5 generator interfaced with PYTHIA6 
using  the  CTEQ6L1 PDF. 
Samples are generated for operators D1, C1, and D9,  
assuming 
$M_*$ = 1~TeV and $m_{\chi}$ between 10~GeV and 1300~GeV. 
Samples for the $b$-FDM model are generated for $m_\chi$ values between 1~GeV and 1300~GeV 
and mediator masses, $m_{\phi}$,  between 5~GeV and 3000~GeV.
The instrumental uncertainties on the simulated signal yields for D1, C1, and D9 operators are between 11\% and 15\%, depending on the signal region. The equivalent uncertainties for the $b$-FDM model range between 6\%-16\% depending on $m_{\chi}$ and the mediator mass. The uncertainties from the PDF are computed by comparing the rates obtained with the default PDF set (CTEQ6L1) with those obtained with two alternative sets (MSTW2008LO and NNPDF21LO \cite{Martin:2009iq, Ball:2011uy}). The uncertainties on the signal acceptance from PDF and scale variations are estimated to be approximately 10\% for the D1, C1, and D9 operators for $m_{\chi}$=10 GeV and approximately $6$\% for  $b$-FDM models.

The validity of the effective field theory assumption depends on the momentum transfer of the process mo\-delled, which should be below the energy scale of the underlying interactions \cite{Busoni:2013lha}. To account for this, the momentum transfer $m(\chi\chi) = Q_\mathrm{tr}$ in the events is required to be less than the energy scale probed. Specifically, $Q_\mathrm{tr}$ must be smaller than the mass $M$ of the heavy mediator. For an ultraviolet completion this implies $M_*=M/\sqrt{g_q g_\chi}$. Along with perturbativity of the couplings $g_q g_\chi<4\pi$ this leads to the following validity requirements on MC truth level: $Q_\mathrm{tr}<4\pi \left( M_*^3/m_q \right)^{1/2}$ (D1), $Q_\mathrm{tr}<4\pi M_*$ (D9), $Q_{\mathrm tr}<(4 \pi)^2 M_*^2 / m_q$ (C1).

\section{Results}

Table~\ref{tab1} shows the expected background from various sources in the four signal regions as well as the observed yields in data.  
The expected signal yields for the operators D1, C1, and D9, as well as for the $b$-FDM model are also shown. 
The probabilities of the background-only hypothesis, $p$-values, for the signal regions SR1, SR2, SR3, and SR4 are 0.09, 0.29, 0.24, and 0.18, respectively.
As no significant excess is observed, limits on the signal yield are set using a 
profile likelihood ratio test following the $CL_s$ prescription~\cite{Read:2002hq}. 
Also given is the 95\% confidence level (CL) upper limit on the number of beyond-the-SM events.
The yields for the  $b$-FDM model are obtained assuming  $m_\chi$=10~GeV and a mediator mass $m_\phi=600$~GeV. 
The limit on $M_*$ for a given assumption on $m_\chi$ is determined by varying $M_*$ and scaling the number of signal events predicted
by the corresponding sample generated with $M_* = 1$ TeV until it is equal to the observed upper limit on beyond-the-SM events.
The corresponding production cross-section for DM produced via the D1 operator in association with $b$($t$)-quarks and $m_\chi=10$~GeV is 38 (221) fb. The cross-section for $b$-FDM models with $m_\phi=600~$~GeV and $m_\chi=10$~GeV is 134 fb.
The signal efficiency is independent of $M_{*}$.

Figure~\ref{DataMC} shows the \met\ distributions for (a) SR1, (b) SR2, and (d) SR4 and (c) the $R$ variable for SR3.

\begin{figure}[ht]
  \centering \begin{tabular}{c@{\hskip 5mm}c}
    \includegraphics[width=.48\textwidth]{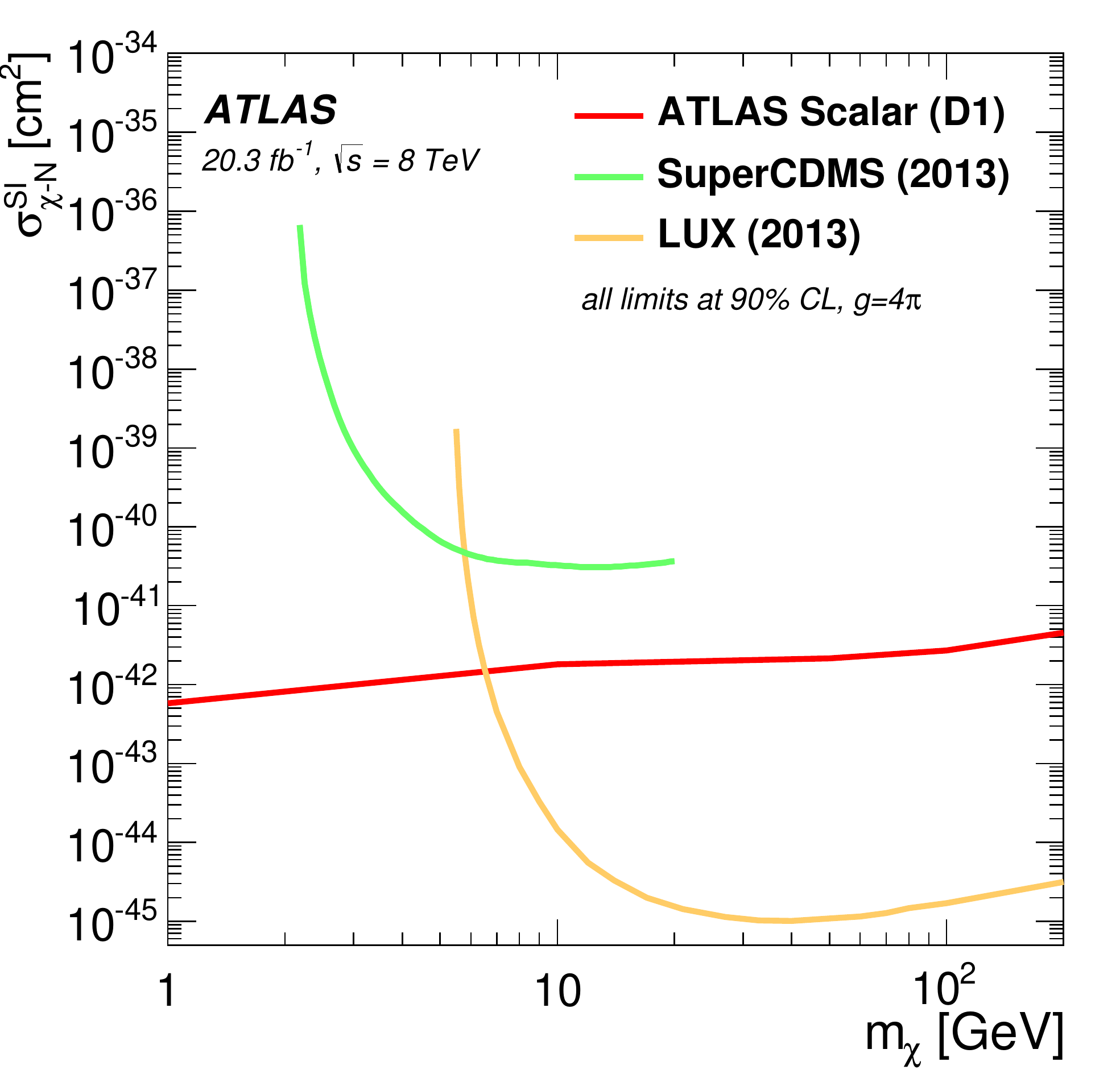}
  \end{tabular} 
  \caption{Upper limits at 90\% CL on the spin-independent $\chi$--nucleon cross-section ($\sigma_\textrm{\tiny $\chi$-N}^{\rm SI}$) 
  for the scalar operator D1 (red)
  as a function of $m_{\chi}$. The yellow and green curves represent the exclusion limits recently set by the LUX and Super-CDMS collaborations~\cite{lux,cdmsSlite,dmtools}. 
  The coupling is assumed to be $g_q g_\chi=g=4 \pi$.
} 
  \label{figxsection_si}
\end{figure}

\begin{figure}[ht]
  \centering \begin{tabular}{c@{\hskip 5mm}c}
    \includegraphics[width=.48\textwidth]{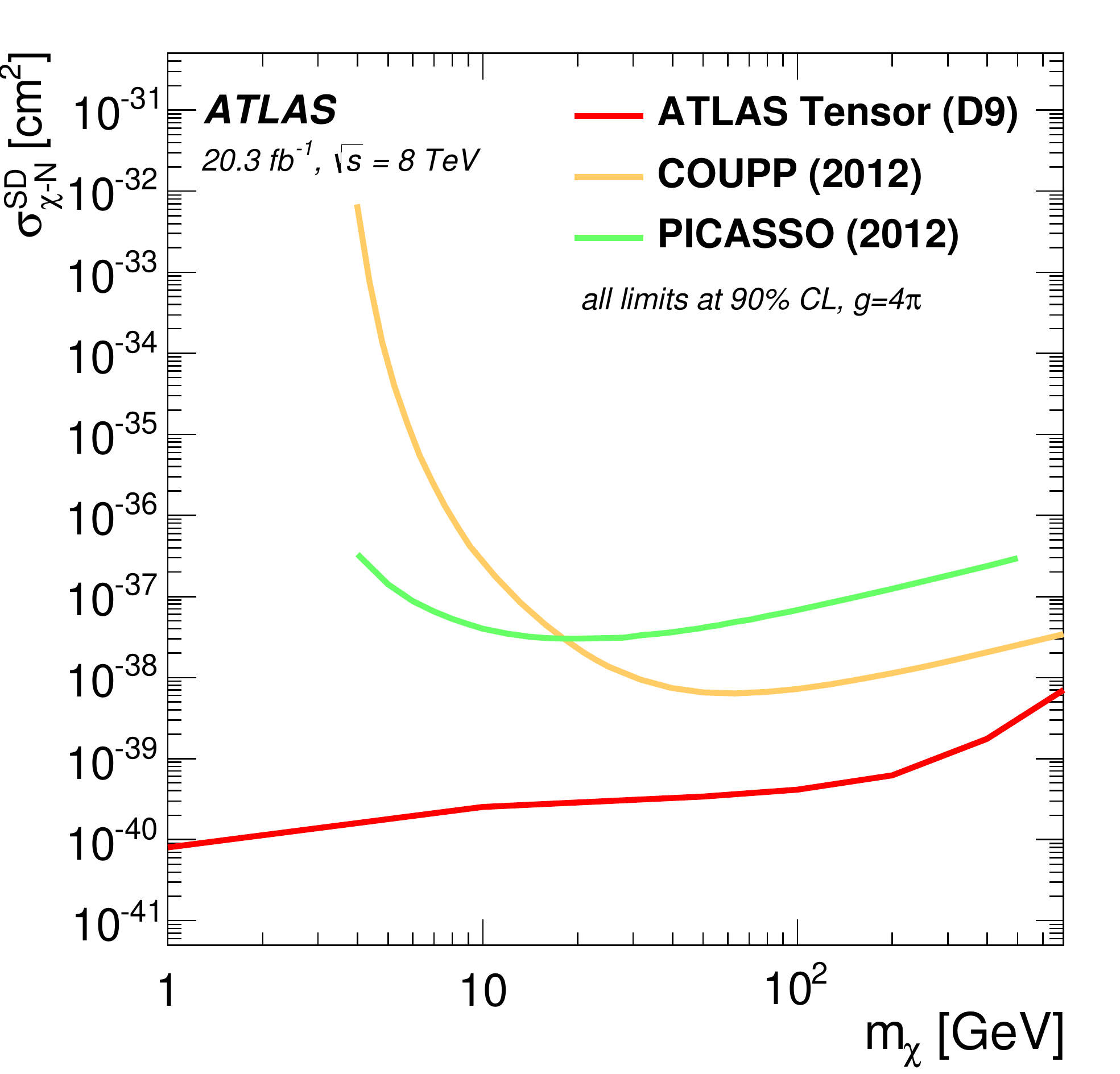}
  \end{tabular} 
  \caption{ Upper limits at 90\% CL on the spin-dependent $\chi$--nucleon cross-section ($\sigma_\textrm{\tiny $\chi$-N}^{\rm SD}$) 
for the tensor operator D9 (red)
  as a function of $m_{\chi}$. The yellow and green curves represent the exclusion limits recently set by the COUPP and PICASSO collaborations~\cite{coupp,picasso,dmtools}. 
  The coupling is assumed to be $g_q g_\chi=g=4 \pi$.
} 
  \label{figxsection_sd}
\end{figure}

\begin{figure}[ht]
    \begin{minipage}{0.48\textwidth}
      \centering 
       \includegraphics[scale=0.45]{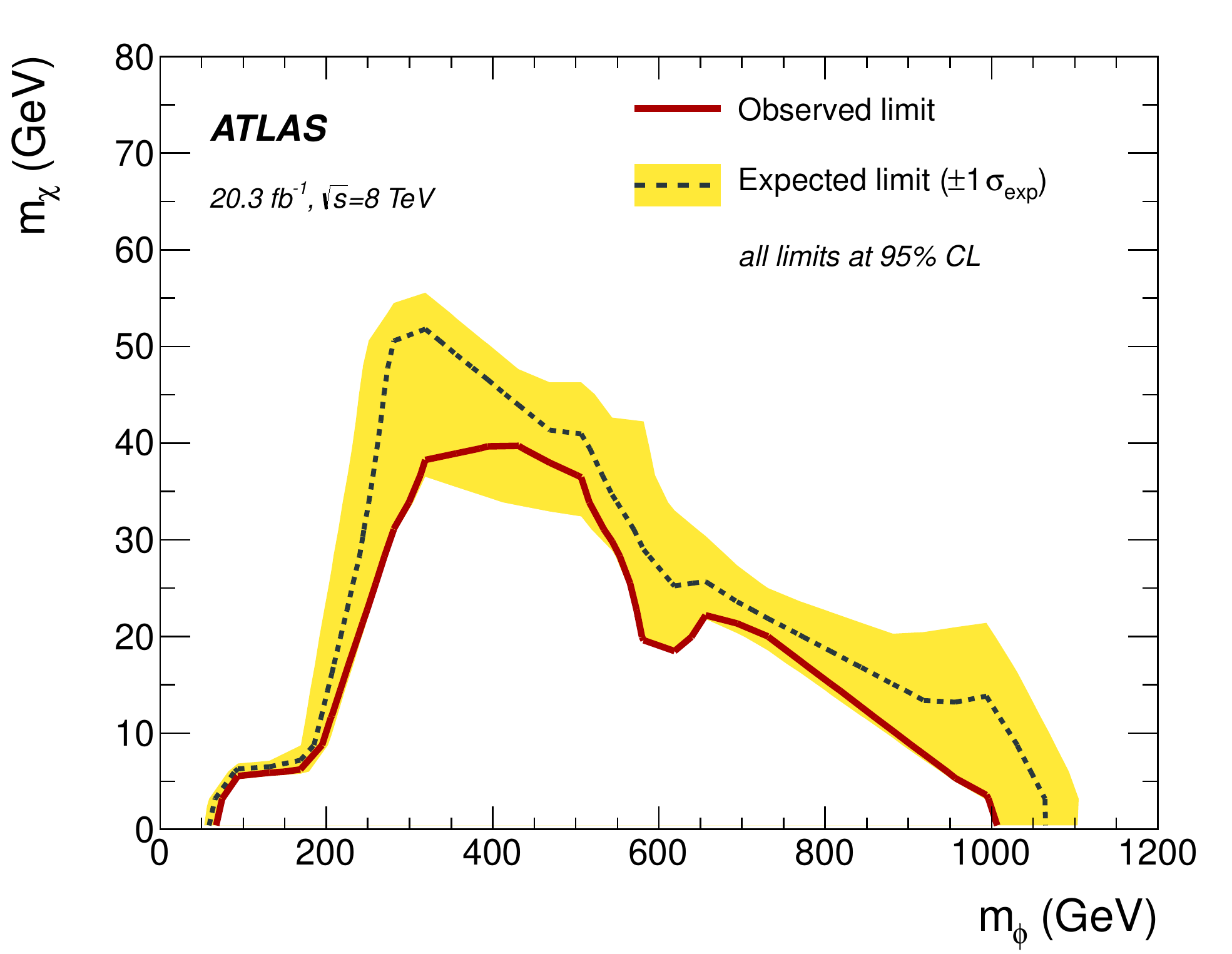}
    \end{minipage}
      \caption{Exclusion contour at 95\% CL for the $b$-FDM model from combined results of SR1 and SR2.  
The expected limit is given by the dashed line, and the yellow band 
      indicates the $\pm1~\sigma$ uncertainty. The observed limit, largely dominated by SR1, is given by the solid red line. The region beneath the curve indicating the observed limit is excluded.
        \label{fig:bfdm_comb_limits2D} }
\end{figure}

%
Figure~\ref{figMstar_90} shows the 90\% CL exclusion curves 
for the effective mass scale $M_*$ as a function of $m_{\chi}$. 
The results for the operators D1, C1, and D9 are presented in\-di\-vi\-du\-ally for all four signal regions. 
The best limits on the D1 and C1 operators are obtained using SR4, while SR1 provides the best limits 
on the D9 operator, as shown in Fig.~\ref{figMstar_90}.
These limits are then converted into limits on the $\chi$--nucleon cross-section~\cite{Goodman:2010ku}. 
Figures~\ref{figxsection_si} and \ref{figxsection_sd} show the corresponding 90\% CL 
exclusion curves for the spin-independent and spin-depen\-dent $\chi$--nucleon cross-section 
for the scalar (D1) and tensor (D9) operators as a function of $m_{\chi}$ for the strongest 
results obtained in any signal region.
The most stringent limits set by direct detection experiments~\cite{lux,cdmsSlite,coupp,picasso}  
are also shown. 
Only $m_\chi$ where more then $90\%$ of the events fulfill the effective field theory validity constraints are shown in Figs.~\ref{figxsection_si} and \ref{figxsection_sd}.
%

The limits shown are especially strong in the low-mass region where several collaborations~\cite{dama,cogent,cdmssi,Daylan:2014rsa}  
have recently claimed  possible observations of DM. 
The  results reported in this article represent the first ATLAS limits on the scalar operator C1 
and they significantly improve  the sensitivity to $\chi$--nucleon interactions mediated by the scalar operator D1
compared to previous ATLAS results~\cite{ATLASmonojet,ATLASmonophoton, ATLASmonoWZ,ATLASmonoZ}.   

Figure~\ref{fig:bfdm_comb_limits2D} shows the exclusion curves observed and expected for the $b$-FDM model 
as a function  of the  mediator and DM masses. For each point in ($m_\chi$, $m_\phi$), 
the signal region with the best expected sensitivity is used, with SR1 dominating over the other signal regions.      
For a DM particle of approximately 35~GeV, as suggested by the interpretation of data recorded by the Fermi-LAT collaboration, mediator masses  between approximately 300~GeV and  500 GeV are excluded at 95\% CL.

 \FloatBarrier 

\section{Conclusions}	


In summary, this article reports a search for dark-matter pair production in association  with bottom or top quarks.
The analysis is performed using \Lumi\ of $pp$ collisions collected  at  $\sqrt{s} = 8$~\TeV{} by the ATLAS detector at the LHC.
The results are interpreted in the framework of an effective field theory to set stringent limits on scalar and tensor interactions 
between Standard Model and DM particles. The data are found to be consistent with the Standard Model expectations, and limits are set on
the mass scale of effective field theories that describe scalar and tensor interactions between
DM and Standard Model particles. The exclusion limits are strongest at low DM masses. 
The limit on the $\chi$--nucleon cross-section mediated by the D1 operator is improved significantly with respect to previously published ATLAS results by obtaining
sensitivities of approximately $\sigma_\textrm{\tiny $\chi$-N}^{\rm SI}=10^{-42}\textrm{~cm}^2$ for $m_\chi=10$~GeV.
Constraints on $b$-Flavoured Dark Matter models, 
suitable to explain a possible signal of annihilating DM, are also presented. 
The excluded regions depend on $m_\chi$ and $m_\phi$. For $m_\chi=35$~GeV, mediator particles with $m_\phi=300$--$500$~GeV are excluded.


\section*{Acknowledgements}

\sloppy 
We would like to thank Tongyan Lin (University of Chicago) for helpful discussions about the models presented and the interplay 
between collider DM constraints and direct and indirect DM experiments.
\input{Acknowledgements-3.tex}
\bibliographystyle{epjc} 
\bibliography{paper}
 
\appendix
\setcounter{figure}{0}
\renewcommand{\figurename}{AUX FIG.}
\onecolumn 
 
\newpage  

\input{appendix.tex}

\clearpage 

\newpage  

\input{atlas_authlist.tex}
\end{document}

%% file: Numbers.tex
\newcommand\etmiss{\ensuremath{E_\mathrm{T}^{\mathrm{miss}}}}

\newcommand\met{\ensuremath{E_\mathrm{T}^{\mathrm{miss}}}}
\newcommand\pt{\ensuremath{p_\mathrm{T}}}

\newcommand\amt{\ensuremath{am_\mathrm{T2}}}
\newcommand\Lumi{\ensuremath{20.3 \mathrm{~fb}^{-1}}}

\newcommand{\Eslash}{\mbox{$ E \kern-0.6em\slash$                }}
\newcommand{\pslash}{\mbox{$ p \kern-0.45em\slash$                }}

\newcommand\sysZnunu{\ensuremath{10}}


%% file: Acknowledgements-3.tex

We thank CERN for the very successful operation of the LHC, as well as the
support staff from our institutions without whom ATLAS could not be
operated efficiently.

We acknowledge the support of ANPCyT, Argentina; YerPhI, Armenia; ARC,
Australia; BMWFW and FWF, Austria; ANAS, Azerbaijan; SSTC, Belarus; CNPq and FAPESP,
Brazil; NSERC, NRC and CFI, Canada; CERN; CONICYT, Chile; CAS, MOST and NSFC,
China; COLCIENCIAS, Colombia; MSMT CR, MPO CR and VSC CR, Czech Republic;
DNRF, DNSRC and Lundbeck Foundation, Denmark; EPLANET, ERC and NSRF, European Union;
IN2P3-CNRS, CEA-DSM/IRFU, France; GNSF, Georgia; BMBF, DFG, HGF, MPG and AvH
Foundation, Germany; GSRT and NSRF, Greece; RGC, Hong Kong SAR, China; ISF, MINERVA, GIF, I-CORE and Benoziyo Center, Israel; INFN, Italy; MEXT and JSPS, Japan; CNRST, Morocco; FOM and NWO, Netherlands; BRF and RCN, Norway; MNiSW and NCN, Poland; GRICES and FCT, Portugal; MNE/IFA, Romania; MES of Russia and NRC KI, Russian Federation; JINR; MSTD,
Serbia; MSSR, Slovakia; ARRS and MIZ\v{S}, Slovenia; DST/NRF, South Africa;
MINECO, Spain; SRC and Wallenberg Foundation, Sweden; SER, SNSF and Cantons of
Bern and Geneva, Switzerland; NSC, Taiwan; TAEK, Turkey; STFC, the Royal
Society and Leverhulme Trust, United Kingdom; DOE and NSF, United States of
America.

The crucial computing support from all WLCG partners is acknowledged
gratefully, in particular from CERN and the ATLAS Tier-1 facilities at
TRIUMF (Canada), NDGF (Denmark, Norway, Sweden), CC-IN2P3 (France),
KIT/GridKA (Germany), INFN-CNAF (Italy), NL-T1 (Netherlands), PIC (Spain),
ASGC (Taiwan), RAL (UK) and BNL (USA) and in the Tier-2 facilities
worldwide.

%% file: appendix.tex
\clearpage 
\newpage 
\section{\Large Auxiliary Material}

\begin{figure*}[htpb]
      \includegraphics[scale=0.29]{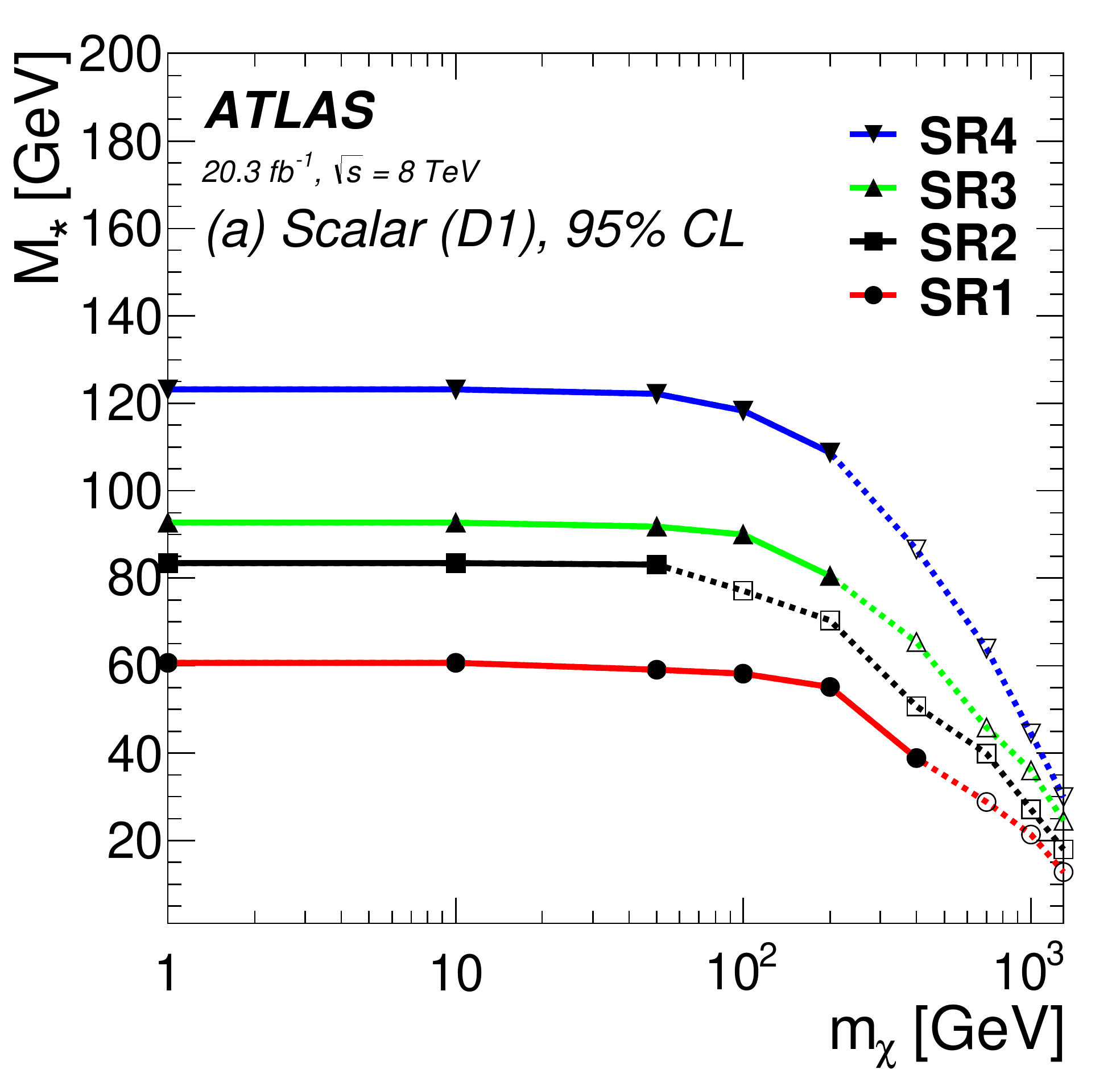}
      \includegraphics[scale=0.29]{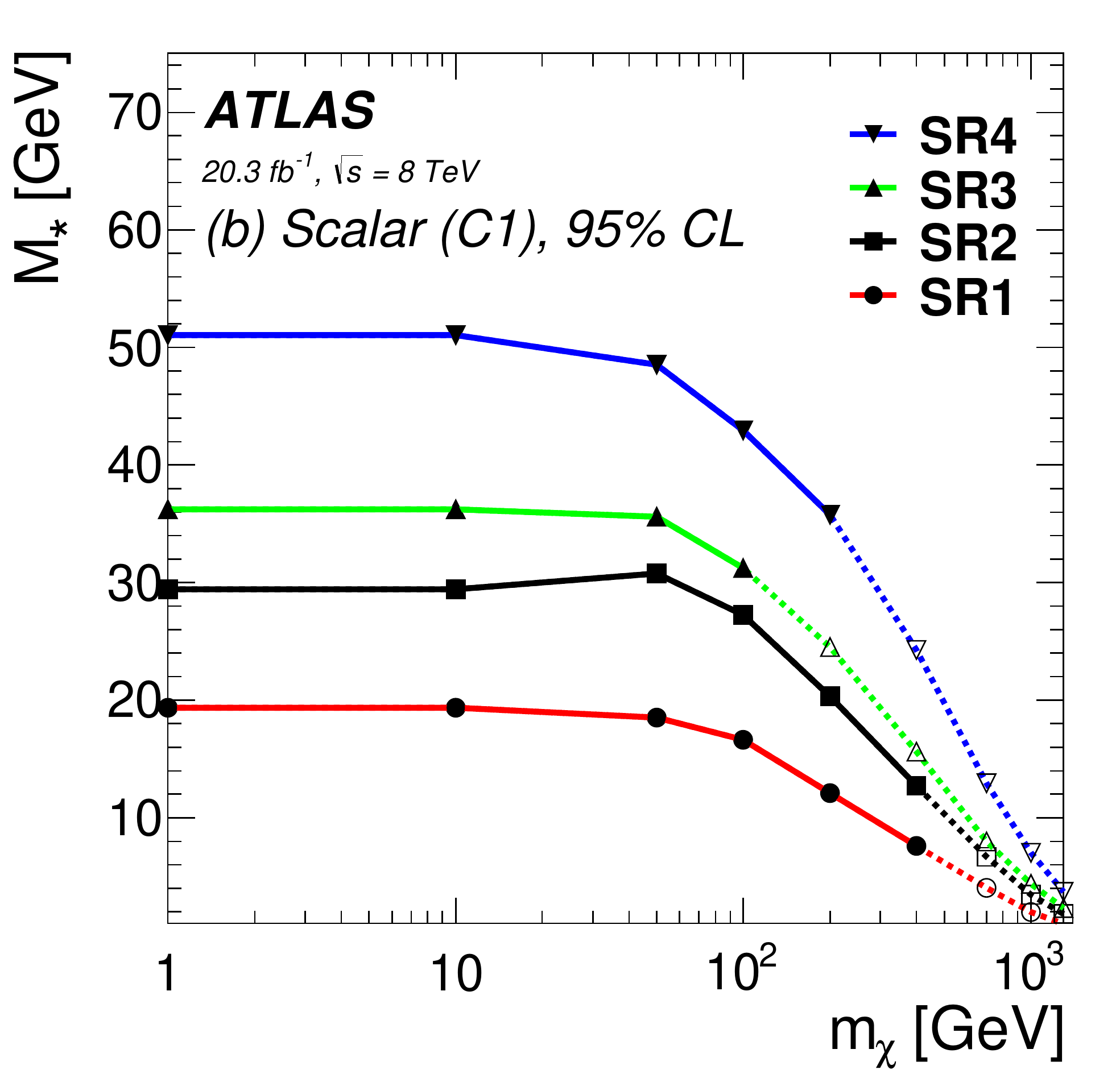}
      \includegraphics[scale=0.29]{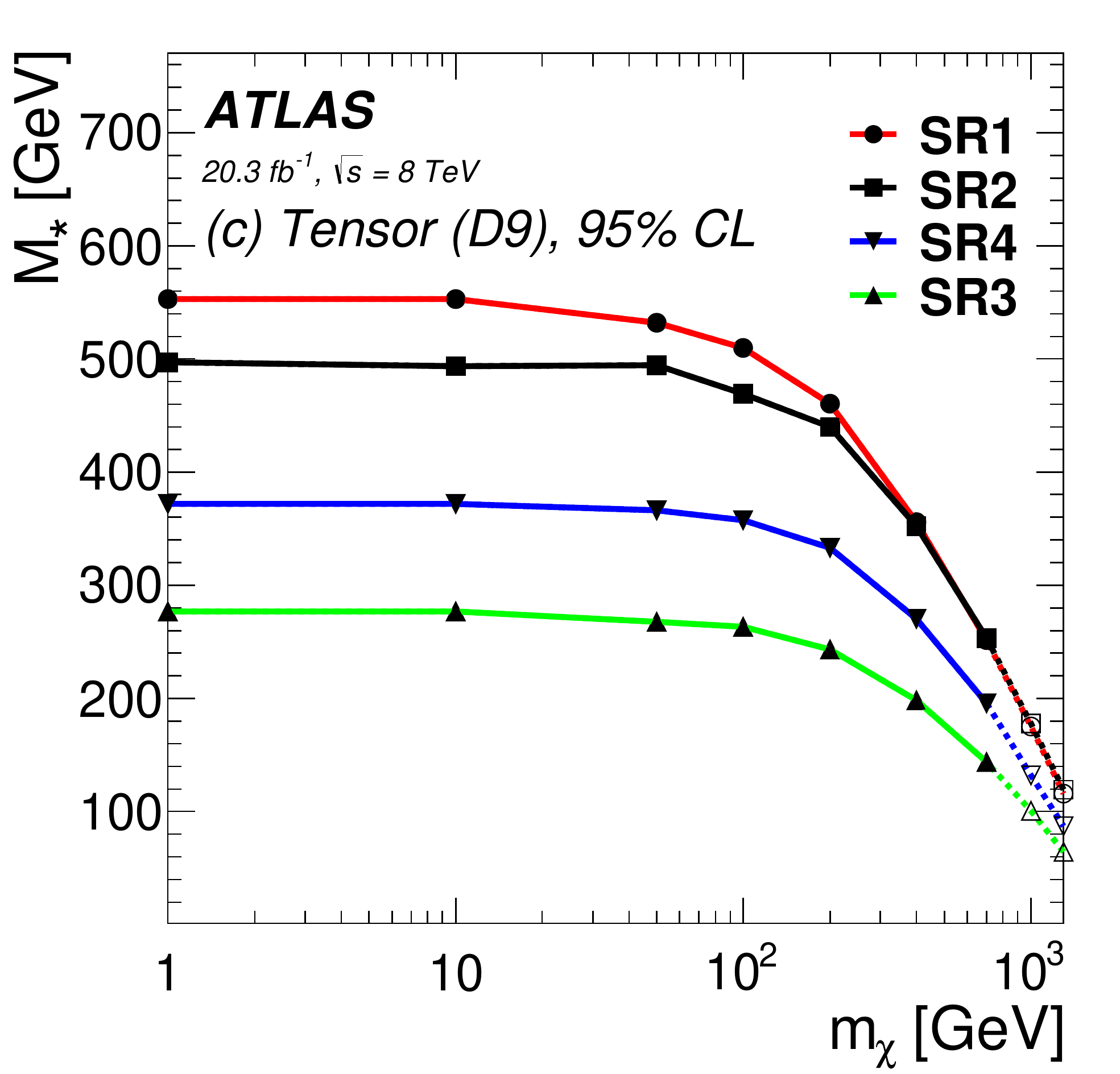}
    \caption{Lower limits on $M_*$ at 95\% CL for the  SR1 (red), SR2 (black), SR3 (green), and SR4 (blue) 
  as a function of $m_{\chi}$ for the operators (a) D1, (b) C1, and (c) D9. Solid lines and markers indicate the validity range of the effective field theory assuming couplings $g_q g_\chi<4\pi$, the dashed lines and hollow markers represent the full collider constraints.}
  \label{figMstar_95}
\end{figure*}

\begin{figure}[h]
\centering
      \includegraphics[scale=0.34]{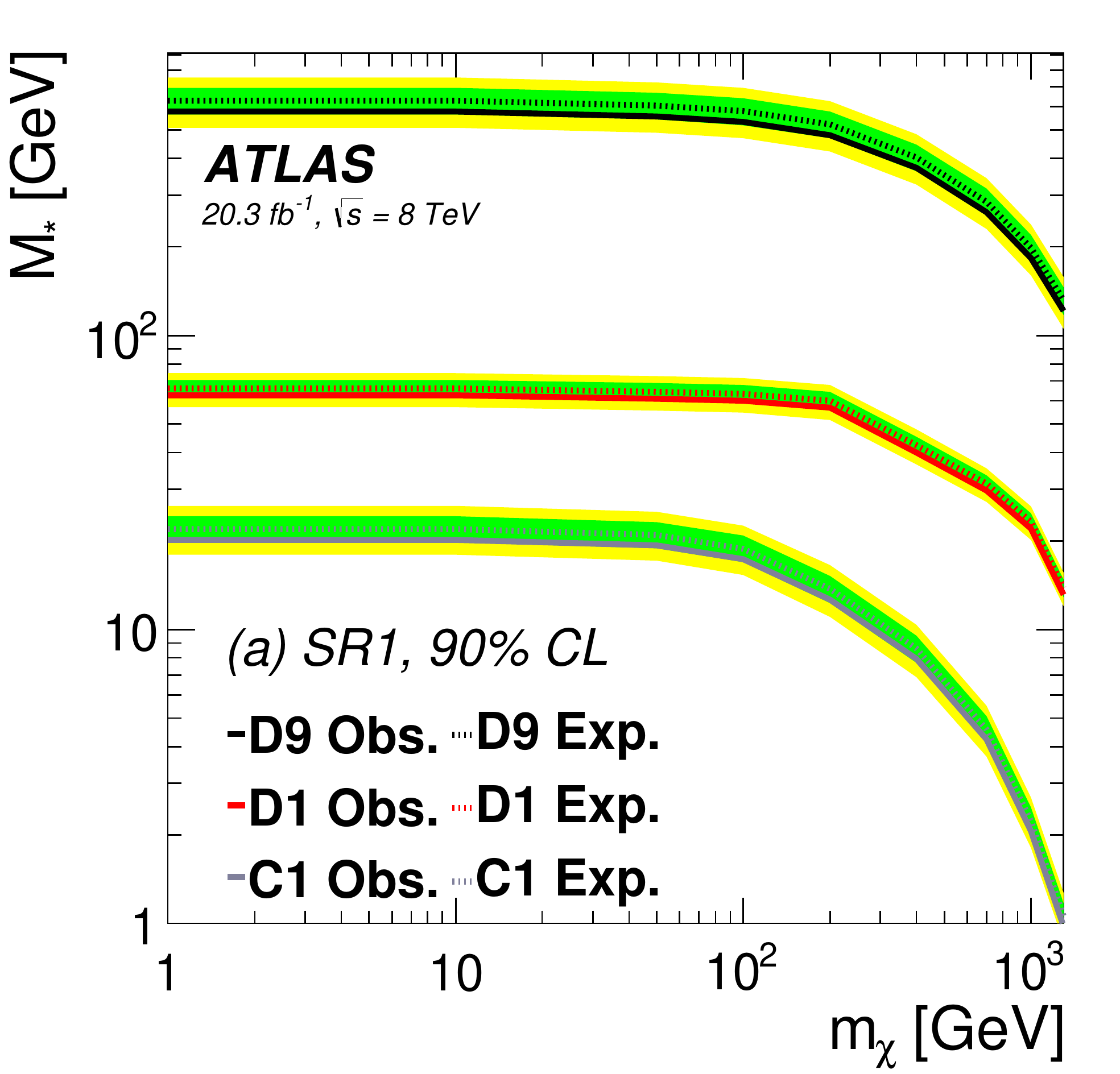}
      \includegraphics[scale=0.34]{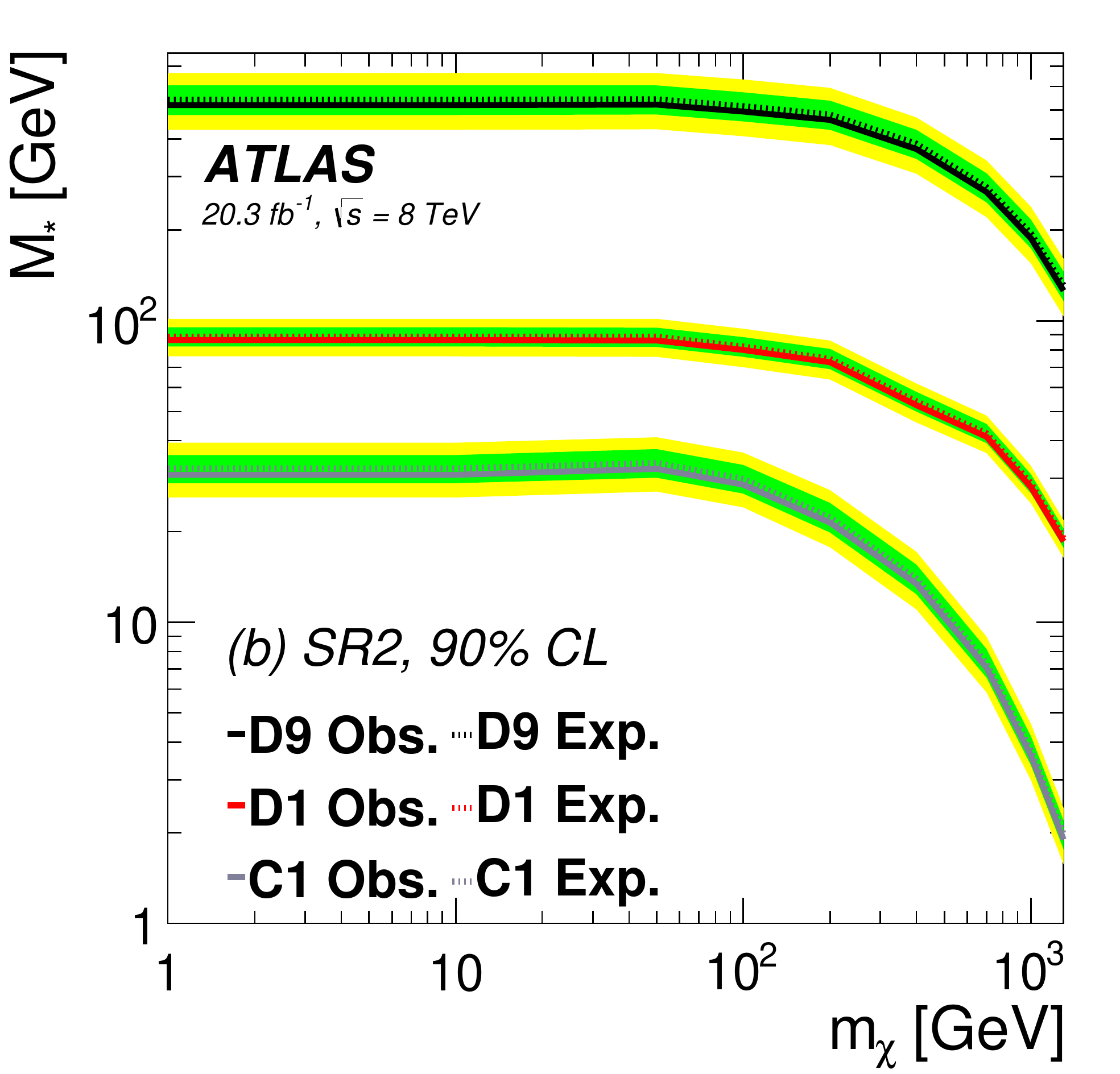}\\
      \includegraphics[scale=0.34]{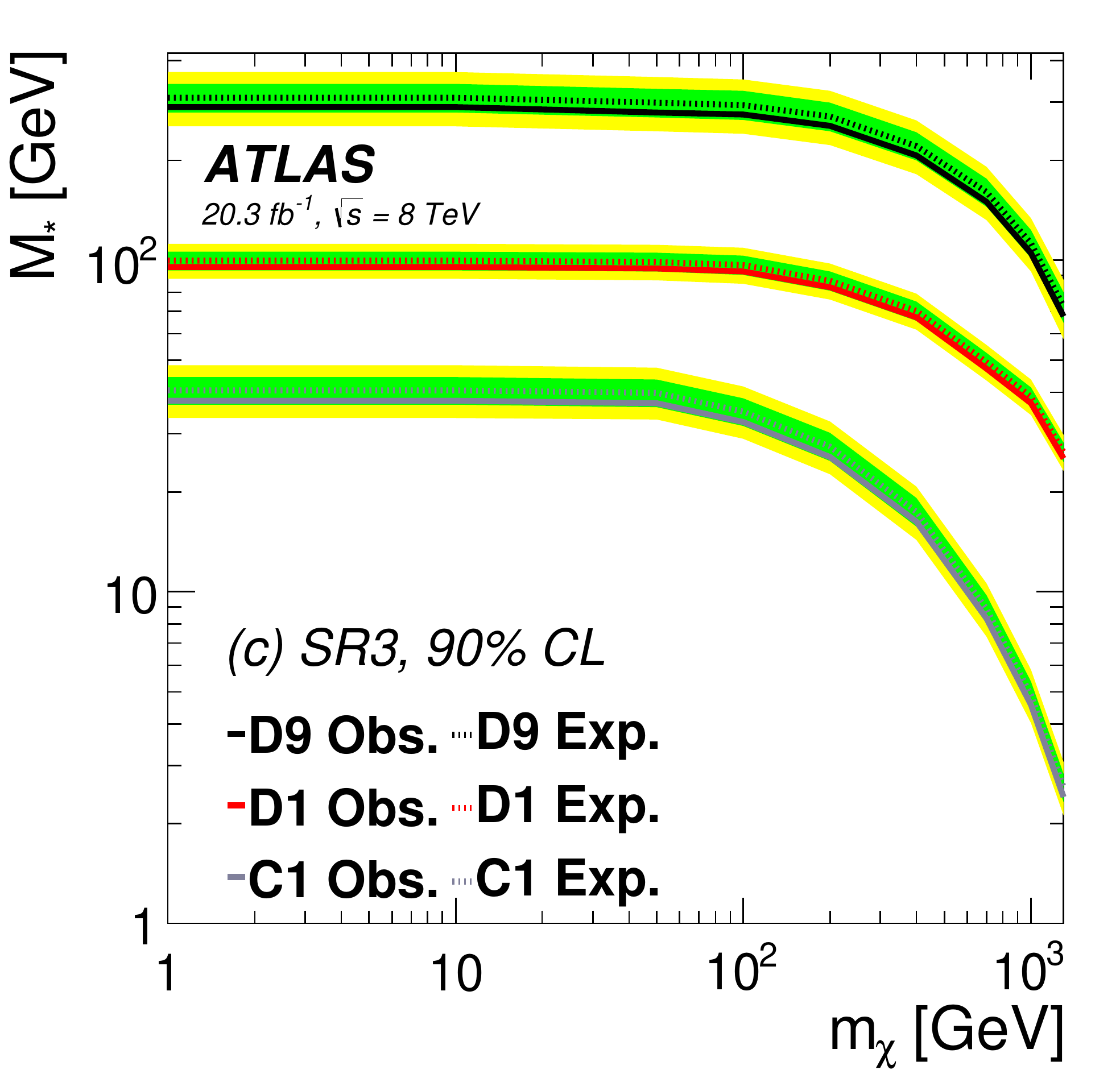}
      \includegraphics[scale=0.34]{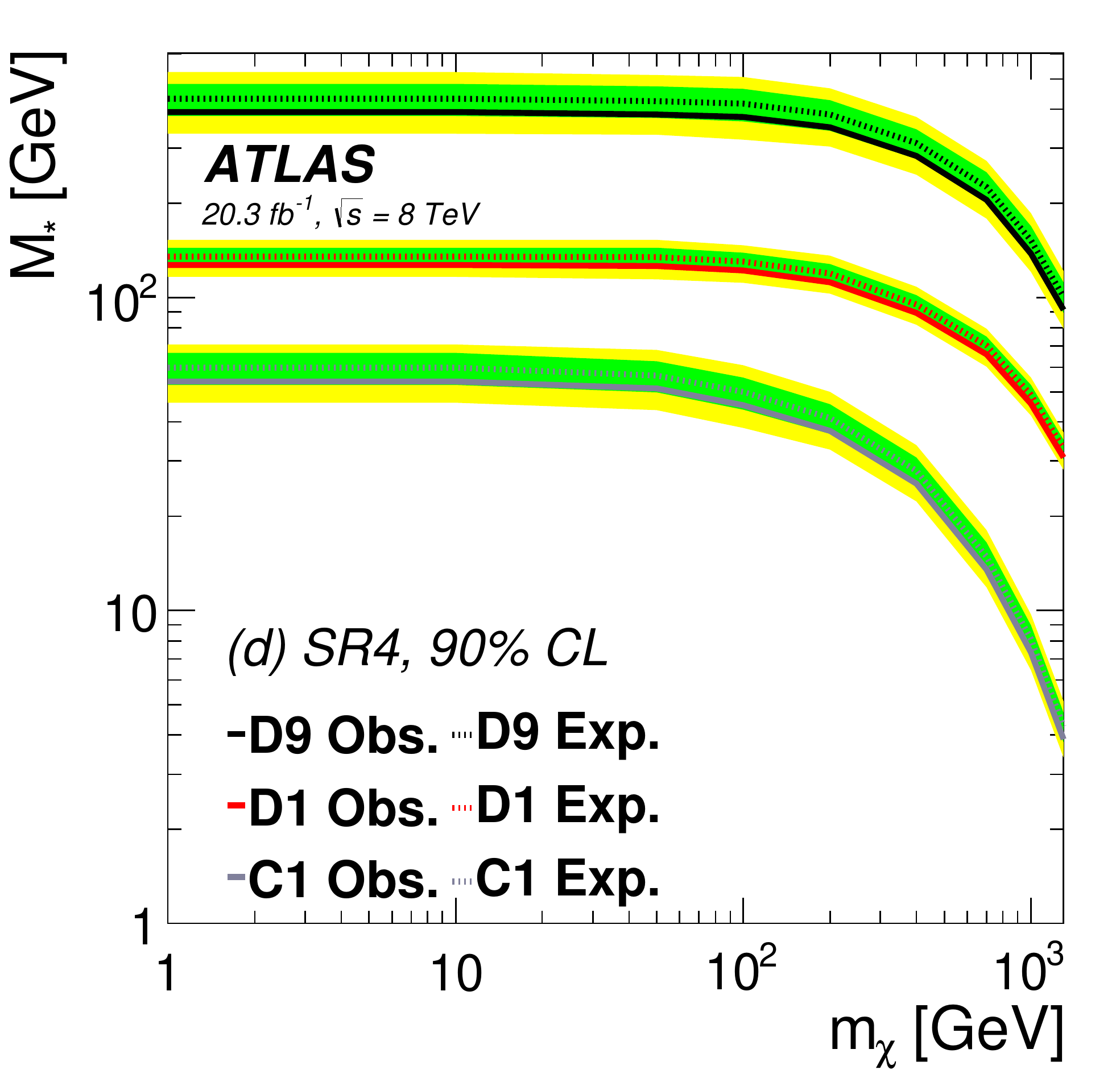}
    \caption{Lower limits on $M_*$ for the  scalar  operators C1 (blue) and D1 (red) and the tensor coupling D9 (black) 
  as a function of $m_{\chi}$ for (a) SR1, (b) SR2, (c) SR3, and d) SR4. The solid lines represent the observed limits.  
  The green (yellow) areas show the 1$\sigma$ (2$\sigma$) regions around 
  the expected limits (dotted lines). }
  \label{figMstar2}
\end{figure}

\begin{figure}[h]
\centering
  \includegraphics[width=.50\textwidth]{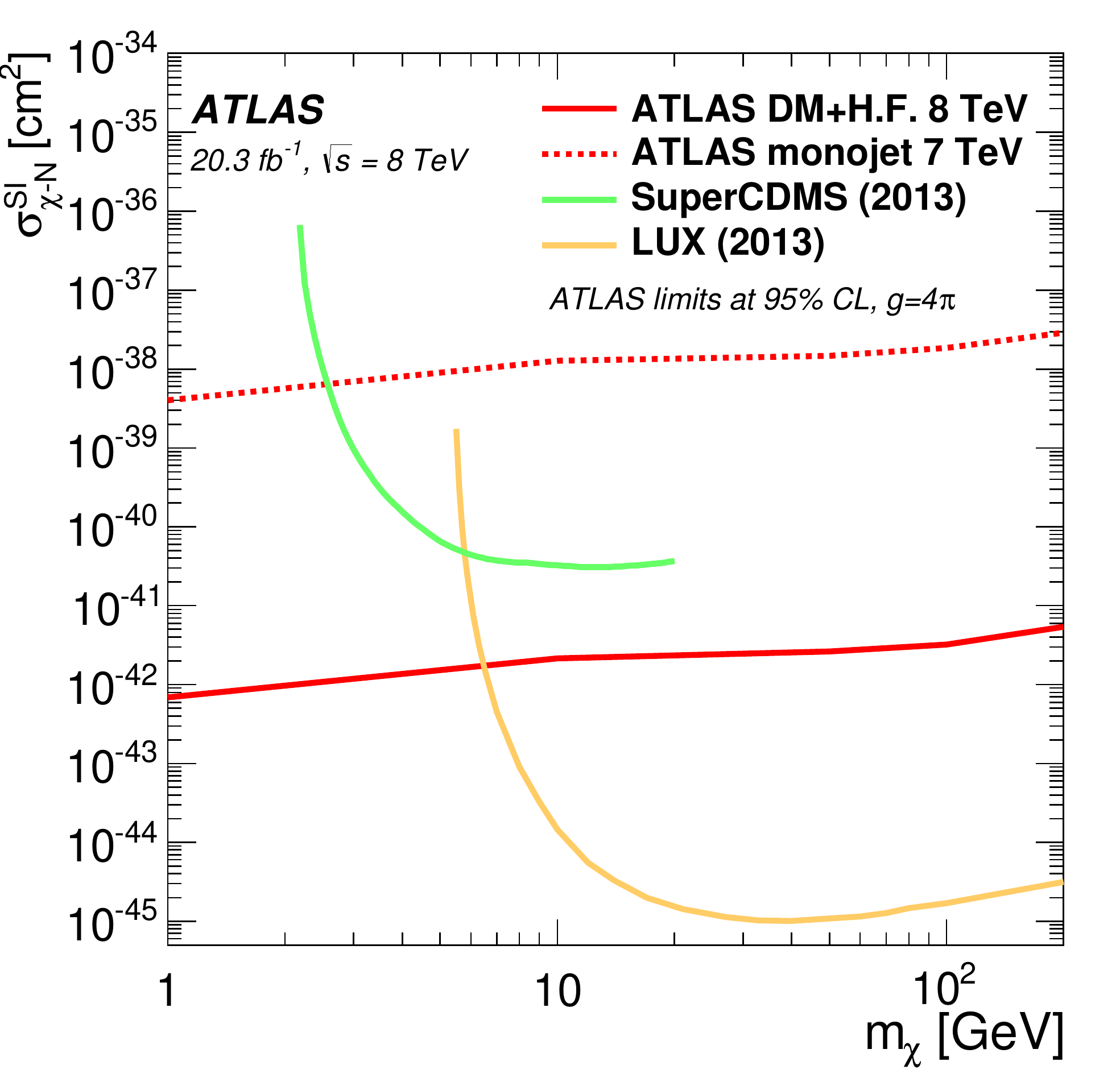}   
  \caption{Comparison between limits on $\sigma_\textrm{\tiny $\chi$-N}$ for the D1 operator 
  obtained in this work and those previously published by ATLAS in~\cite{ATLASmonojet}.  
  ATLAS limits are set at 95\% CL The ATLAS curves shown correspond to the D1 operator.
  The coupling is assumed to be $g_q g_\chi=g=4 \pi$.
 }
  \label{comparison-D1-ATLASmonojet}
\end{figure}

\begin{figure}[h]
\centering
  \includegraphics[width=.50\textwidth]{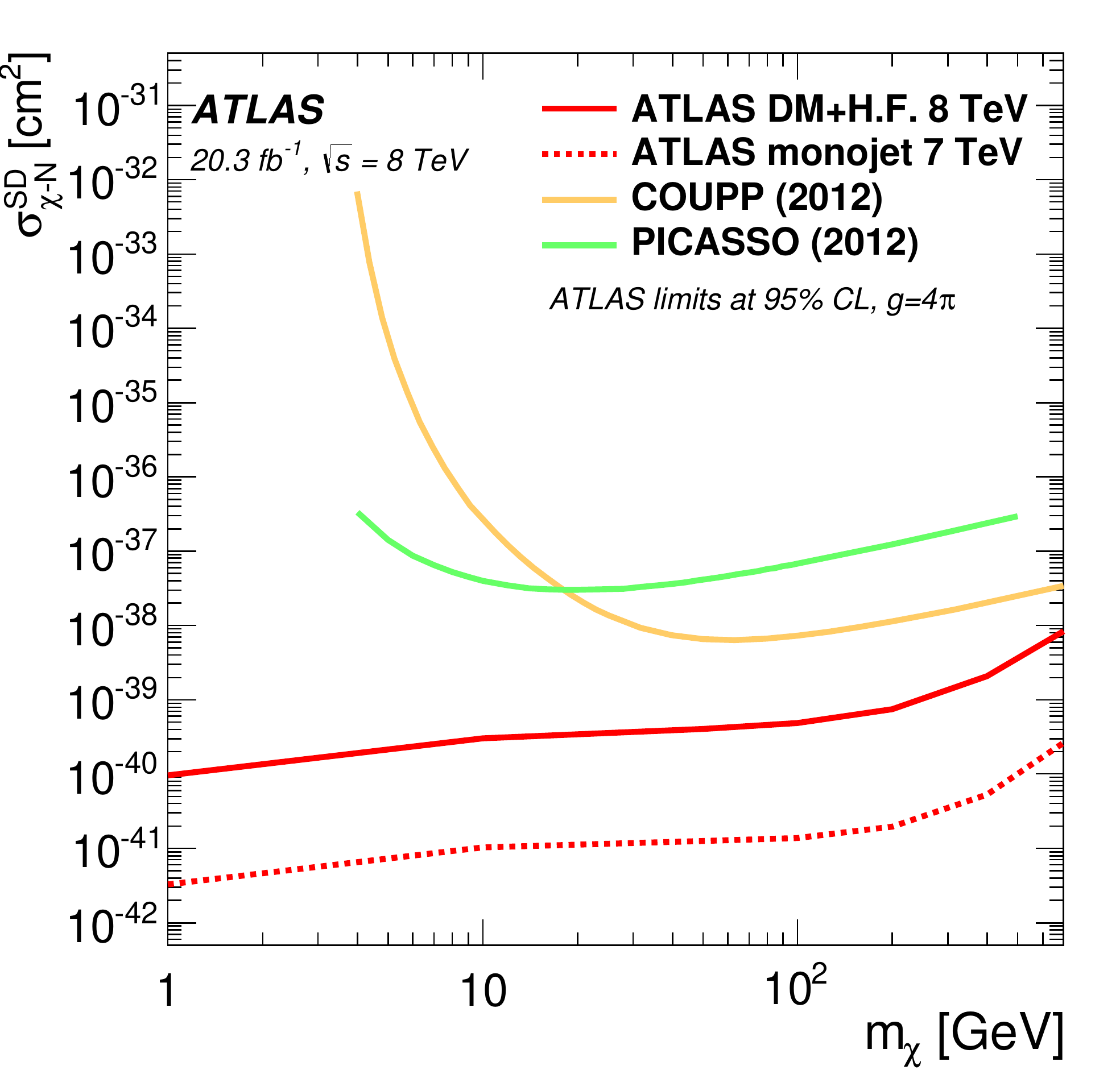}   
  \caption{Comparison between limits on $\sigma_\textrm{\tiny $\chi$-N}$ for the D9 operator 
  obtained in this work and those previously published by ATLAS in~\cite{ATLASmonojet}. 
  ATLAS limits are set at 95\% CL The ATLAS curves shown correspond to the D9 operator.  
   The coupling is assumed to be $g_q g_\chi=g=4 \pi$.
}
  \label{comparison-D9-ATLASmonojet}
\end{figure}

\begin{figure}[h]
\centering
  \includegraphics[width=.50\textwidth]{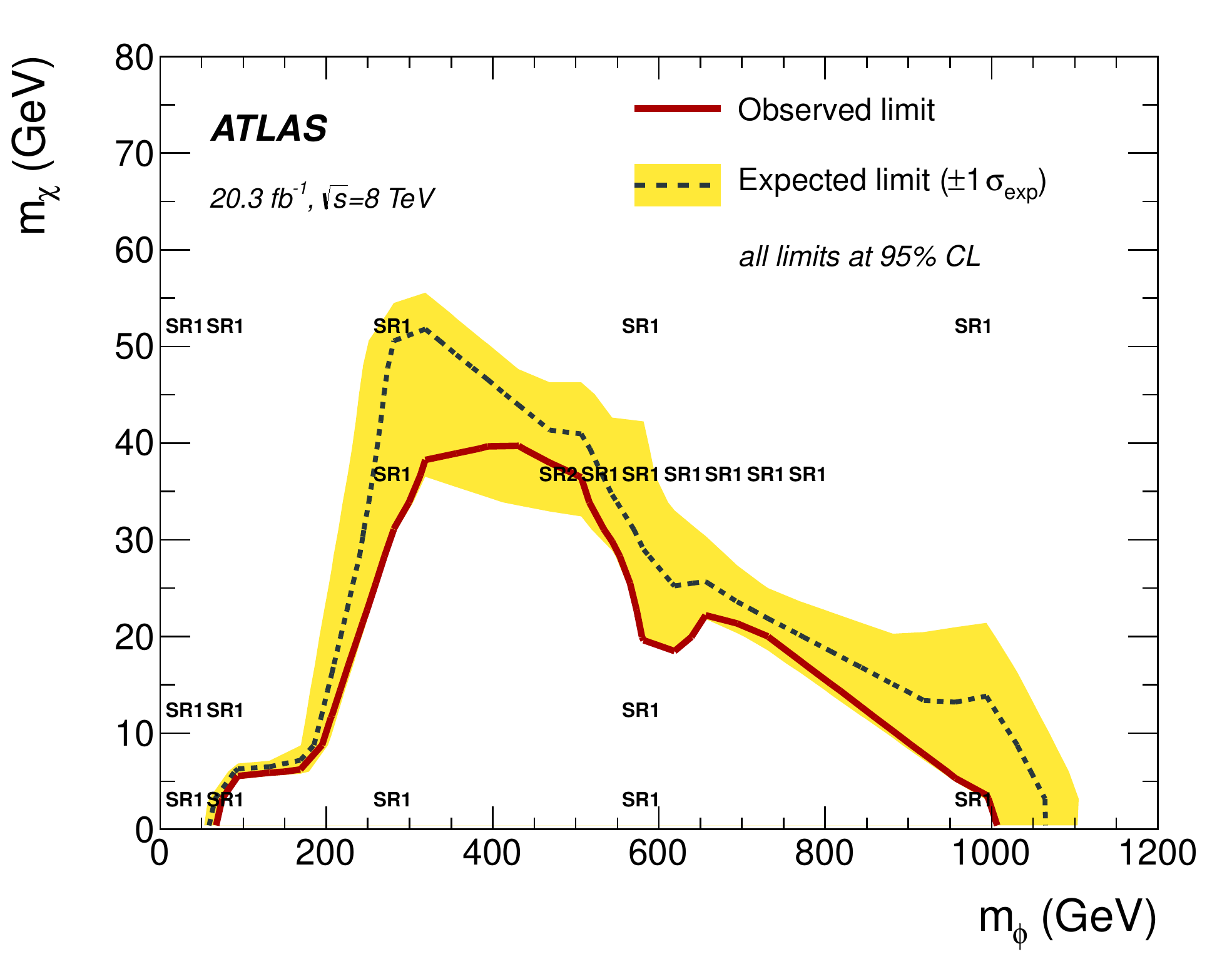}   
  \caption{Exclusion contour for the $b$-FDM model. The expected limit is given by the dashed line, and the yellow band indicates the $\pm1~\sigma$ uncertainty. The observed limit is given by the solid red line. The region beneath the observed curve is excluded. The signal region (SR) leading to best expected results are indicated for each DM mass and operator.
  }
  \label{bestlimits-bFDM}
\end{figure}


\begin{table*}[h]
  \centering
  \caption{
  Expected number of signal events for selected signal samples as a function of $m_\chi$ and  $m_\phi$, if applicable. The uncertainties correspond to statistical and systematic uncertainties added in quadrature. The mass scale $M_*$ is set to be 100/40/600 GeV for the D1/C1/D9 operators, corresponding to approximately the expected limit.
 }          
  \label{tab:signal_yields}
  \newcommand{\twrw}[1]{\multirow{2}{*}{#1}}
    \begin{tabular}{lcc r@{$\,\pm \,$}l r@{$\,\pm \,$}l r@{$\,\pm \,$}l r@{$\,\pm \,$}l r@{$\,\pm \,$}l r@{$\,\pm \,$}l} \hline \hline
  Signal sample              & \multicolumn{2}{c}{$m_\chi$ [GeV]} &     \multicolumn{2}{c}{SR1} & \multicolumn{2}{c}{SR2} & \multicolumn{2}{c}{SR3} & \multicolumn{2}{c}{SR4}\\\hline\hline
  Expected signal -- D1      & \multicolumn{2}{c}{100}      & 8 & 1       &   32 & 6      &    24 & 2     &  28 & 4 \\
  Expected signal -- D1      & \multicolumn{2}{c}{200}      & 6 & 1       &   18 & 3      &    12 & 1     &  17 & 2 \\
  Expected signal -- D1      & \multicolumn{2}{c}{700}      & 0.12 & 0.02 &   0.6 & 0.1   &  0.42 & 0.03 &  0.7 & 0.1 \\\hline

  Expected signal -- C1      & \multicolumn{2}{c}{100}      &  9 & 1      &   45   & 7     &   25 & 2     &  24   & 6 \\
  Expected signal -- C1      & \multicolumn{2}{c}{200}      &  2.8 & 0.5  &   15   & 3     &    9 & 1     &  11   & 2 \\
  Expected signal -- C1      & \multicolumn{2}{c}{700}      & 0.04 & 0.01 &   0.19 & 0.03  & 0.12 & 0.01  &  0.18 & 0.02 \\\hline

  Expected signal -- D9      & \multicolumn{2}{c}{100}      &  107 & 17   &   57 & 10    &   1.8 & 0.2  &   1.5 &   0.4 \\
  Expected signal -- D9      & \multicolumn{2}{c}{200}      &  72  & 11   &   44 & 7     &   1.3 & 0.1  &   1.0 &   0.2 \\
  Expected signal -- D9      & \multicolumn{2}{c}{700}      &  7   & 1    &   5  & 1     &  0.16 & 0.01 &  0.12 &   0.01          \\\hline \hline

                             & $m_\chi$ [GeV] & $m_\phi$ [GeV]  &    \multicolumn{2}{c}{SR1} & \multicolumn{2}{c}{SR2} & \multicolumn{2}{c}{SR3} & \multicolumn{2}{c}{SR4}\\\hline
  Expected signal -- $b$-FDM & 35        & 300       &  168 & 23  &   95 & 14    & \multicolumn{2}{c}{--} & \multicolumn{2}{c}{--}\\ 
  Expected signal -- $b$-FDM & 35        & 600       &   83 & 10  &   44 & 5     & \multicolumn{2}{c}{--} & \multicolumn{2}{c}{--}\\ 
  Expected signal -- $b$-FDM & 200       & 600       &   36 & 4   &   31 & 4     & \multicolumn{2}{c}{--} & \multicolumn{2}{c}{--}\\ 
  Expected signal -- $b$-FDM & 1         & 600       &  964 & 122 &  288 & 40    & \multicolumn{2}{c}{--} & \multicolumn{2}{c}{--}\\ \hline \hline

   \end{tabular}
\end{table*}

\begin{table*}
  \centering
  \caption{Product of detector acceptance A and reconstruction efficiency $\epsilon$ (A$\times \epsilon$) for selected signal samples. Acceptances and efficiencies are independent of the mass scale $M_*$.}
  \label{tab:acceff}
  \begin{tabular}{lcc r@{}l r@{}l r@{}l r@{}l}\hline\hline
  Signal sample                          & \multicolumn{2}{c}{$m_\chi$  [GeV]}&   \multicolumn{2}{c}{SR1} & \multicolumn{2}{c}{SR2} & \multicolumn{2}{c}{SR3} &\multicolumn{2}{c}{SR4}       \\\hline\hline
  D1      & \multicolumn{2}{c}{100}      &   0 & .26\% &  1 & .0\%    &  0 & .91\%   & 1 & .1\%  \\
  D1      & \multicolumn{2}{c}{200}      &   0 & .26\% &  1 & .1\%    &  0 & .83\%   & 1 & .1\%  \\
  D1      & \multicolumn{2}{c}{700}      &   0 & .26\% &  1 & .4\%    &  1 & .0\%    & 1 & .7\%  \\\hline

  C1      & \multicolumn{2}{c}{100}      &   0 & .19\% &  0 & .90\%    &  0 & .60\%   & 0 & .60\%  \\
  C1      & \multicolumn{2}{c}{200}      &   0 & .20\% &  1 & .1\%     &  0 & .78\%   & 0 & .90\%  \\
  C1      & \multicolumn{2}{c}{700}      &   0 & .28\% &  1 & .5\%     &  0 & .99\%   & 1 & .6\%   \\\hline

  D9      & \multicolumn{2}{c}{100}      &   3 & .7\% &  2 & .0\%    &  2 & .0\%   & 1 & .8\% \\
  D9      & \multicolumn{2}{c}{200}      &   5 & .5\% &  3 & .3\%    &  2 & .2\%   & 1 & .9\% \\
  D9      & \multicolumn{2}{c}{700}      &   10&.0\%  &  7 & .6\%    &  2 & .7\%   & 2 & .2\% \\\hline \hline

                      & $m_\chi$ [GeV]    & $m_\phi$ [GeV]  &   \multicolumn{2}{c}{SR1} & \multicolumn{2}{c}{SR2} & \multicolumn{2}{c}{SR3} &\multicolumn{2}{c}{SR4}\\ \hline
  $b$-FDM & 35        & 300       & 1  & .6\%  &  0  & .90\%   &   \multicolumn{2}{c}{--}   &       \multicolumn{2}{c}{--}        \\ 
  $b$-FDM & 35        & 600       & 11 & .0\%  &  5  & .7\%    &   \multicolumn{2}{c}{--}   &       \multicolumn{2}{c}{--}        \\ 
  $b$-FDM & 200       & 600       & 15 & .0\%  &  14 &.0\%     &   \multicolumn{2}{c}{--}   &       \multicolumn{2}{c}{--}        \\ 
  $b$-FDM & 1         & 600       & 4  & .1\%  &  1  & .2\%    &   \multicolumn{2}{c}{--}   &       \multicolumn{2}{c}{--}        \\ \hline \hline

   \end{tabular}
\end{table*}


%
%
%
%
%

\begin{table*}
  \centering
  \caption{Production cross sections $\sigma$ for selected samples. The cross sections for the effective field theory operators correspond to the coupling strengths probed. }
  \label{tab:xsec}
  \begin{tabular}{lcc r@{}l r@{}l }\hline\hline
  Signal sample            & \multicolumn{2}{c}{$m_\chi$ [GeV]}  & \multicolumn{2}{c}{$\sigma (pp\to \chi\bar{\chi}+b(\bar{b}))$ [pb]} & \multicolumn{2}{c}{$ \sigma (pp\to \chi\bar{\chi}+t\bar{t})$ [pb]}   \\\hline\hline
  D1      & \multicolumn{2}{c}{100}      &  0&.022   & 0&.13    \\
  D1      & \multicolumn{2}{c}{200}      &  0&.009   & 0&.073    \\
  D1      & \multicolumn{2}{c}{700}      &  0&.00010  & 0&.002    \\\hline
 
  C1      & \multicolumn{2}{c}{100}      &  0&.048   & 0&.20    \\
  C1      & \multicolumn{2}{c}{200}      &  0&.011   & 0&.060    \\
  C1      & \multicolumn{2}{c}{700}      &  0&.000042 & 0&.00057    \\\hline

  D9      & \multicolumn{2}{c}{100}      &  0&.14    & 0&.0044    \\
  D9      & \multicolumn{2}{c}{200}      &  0&.063   & 0&.0028    \\
  D9      & \multicolumn{2}{c}{700}      &  0&.003   & 0&.00028   \\\hline\hline

           & $m_\chi$ [GeV] & $m_\phi$ [GeV] & \multicolumn{4}{c}{$\sigma$ [pb]} \\\hline
  $b$-FDM & 35        & 300       & \multicolumn{4}{c}{0.53}      \\ 
  $b$-FDM & 35        & 600       & \multicolumn{4}{c}{0.039}      \\ 
  $b$-FDM & 200       & 600       & \multicolumn{4}{c}{0.012}      \\ 
  $b$-FDM & 1         & 600       & \multicolumn{4}{c}{1.2}      \\ \hline \hline

   \end{tabular}
\end{table*}

\begin{figure}[h]
\centering
      \includegraphics[scale=0.34]{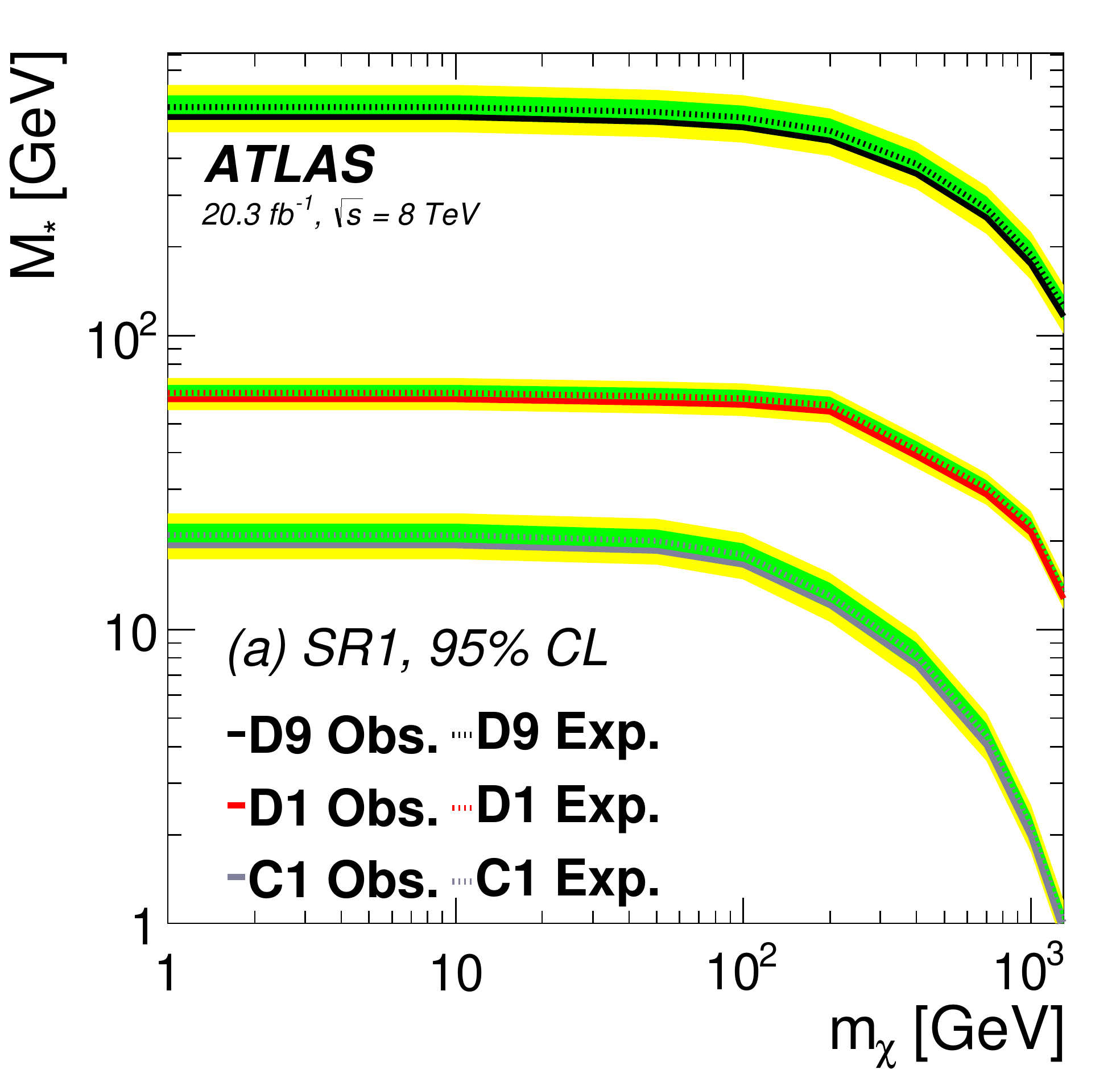}
      \includegraphics[scale=0.34]{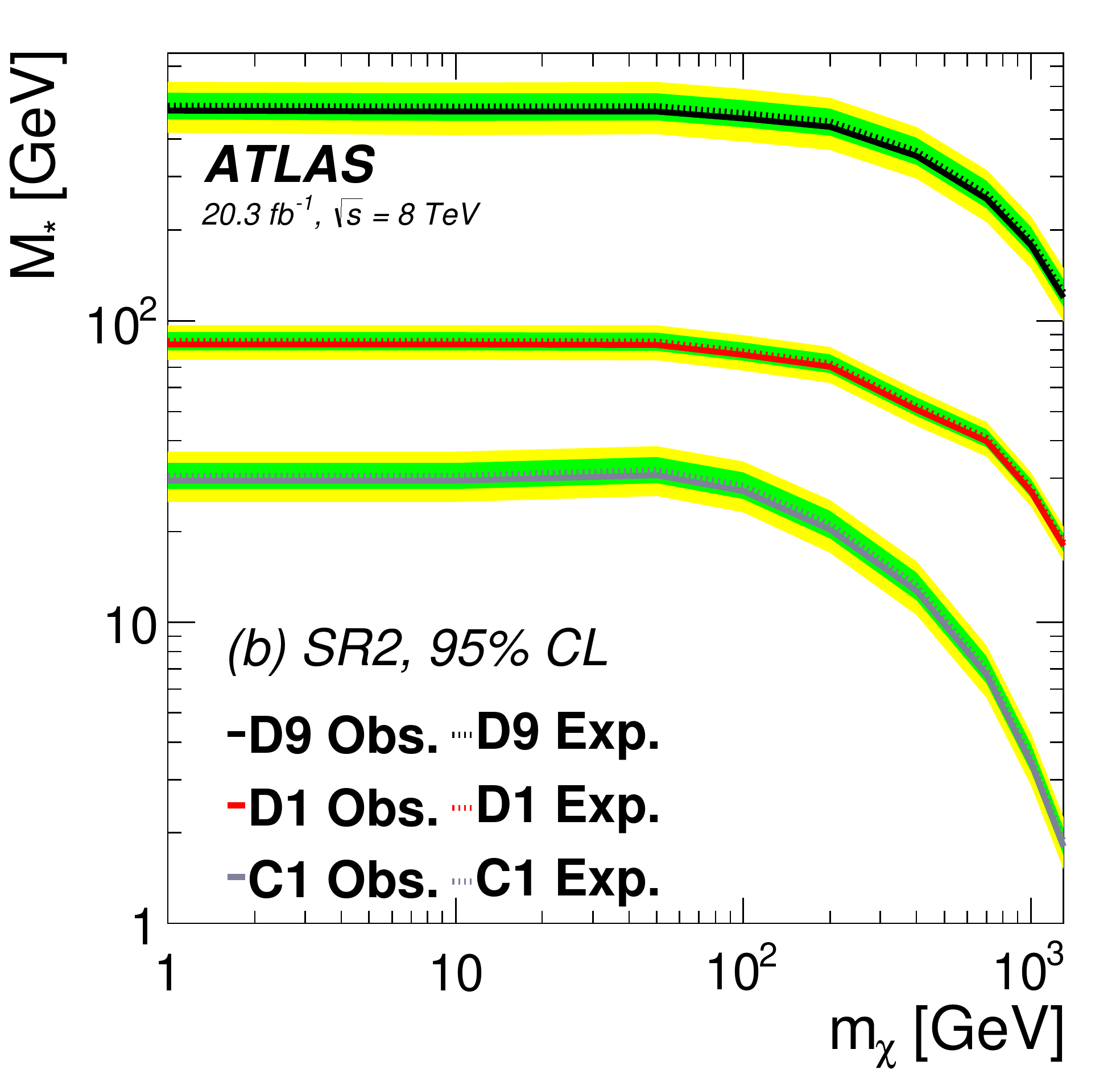}\\
      \includegraphics[scale=0.34]{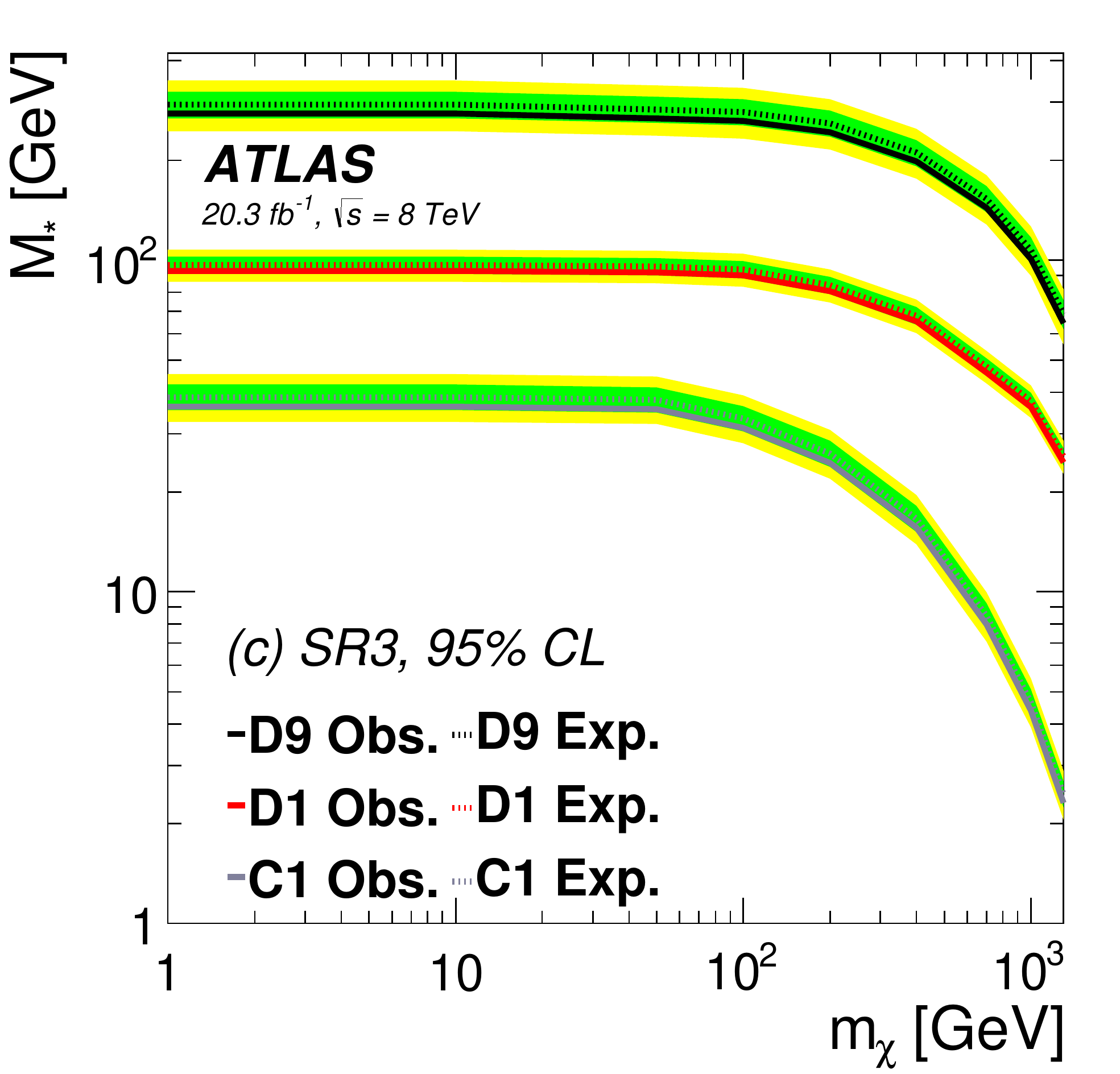}
      \includegraphics[scale=0.34]{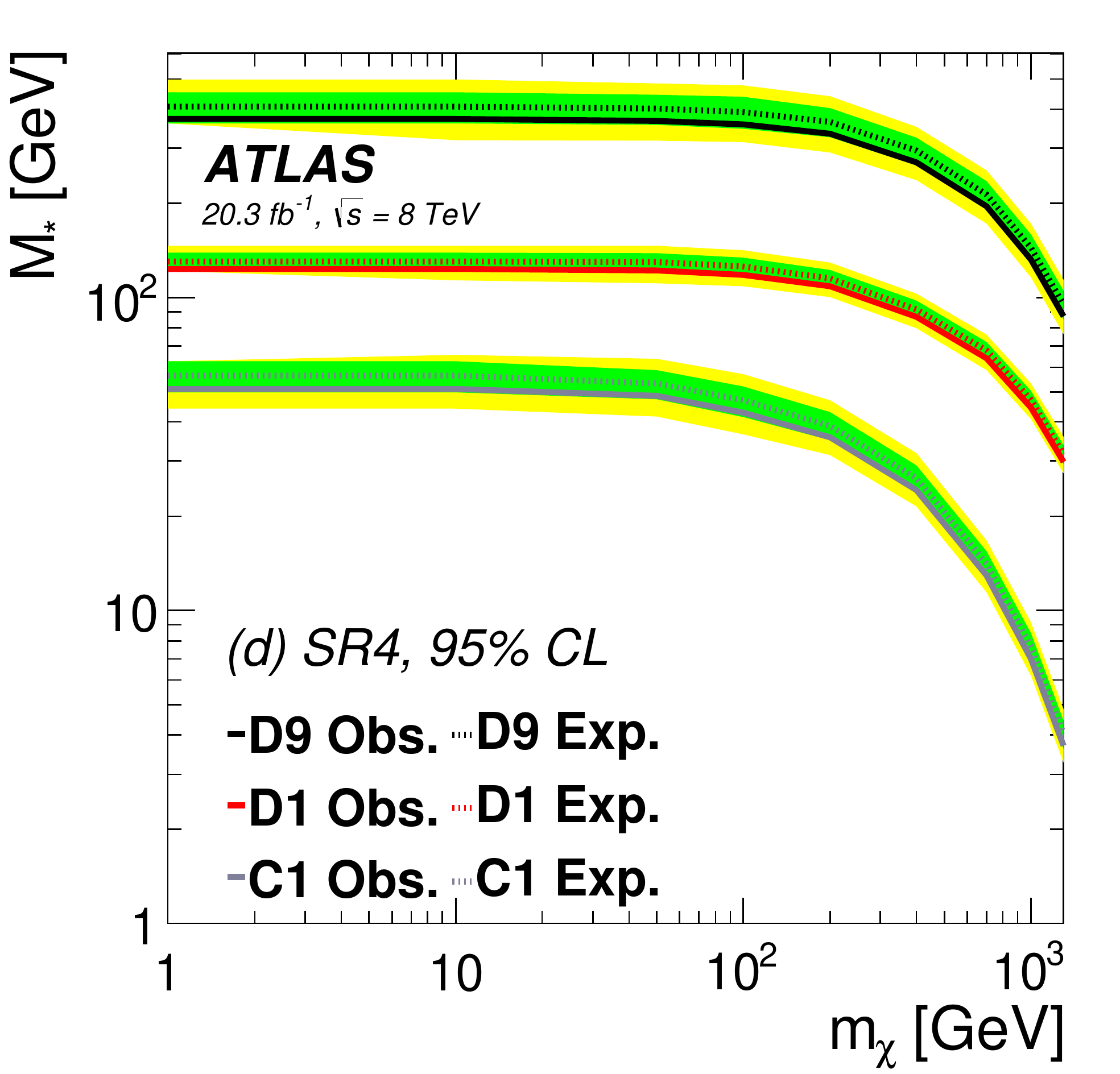}
    \caption{Lower limits on $M_*$ at 95\% CL for the  scalar  operators C1 (blue) and D1 (red) and the tensor coupling D9 (black) 
  as a function of $m_{\chi}$ for (a) SR1, (b) SR2, (c) SR3, and (d) SR4. The solid lines represent the observed limits.  
  The green (yellow) areas show the 1$\sigma$ (2$\sigma$) regions around 
  the expected limits (dotted lines). }
  \label{figMstar2_95}
\end{figure}


\begin{figure}[h]
  \centering \begin{tabular}{c@{\hskip 5mm}c}
    \includegraphics[width=.5\textwidth]{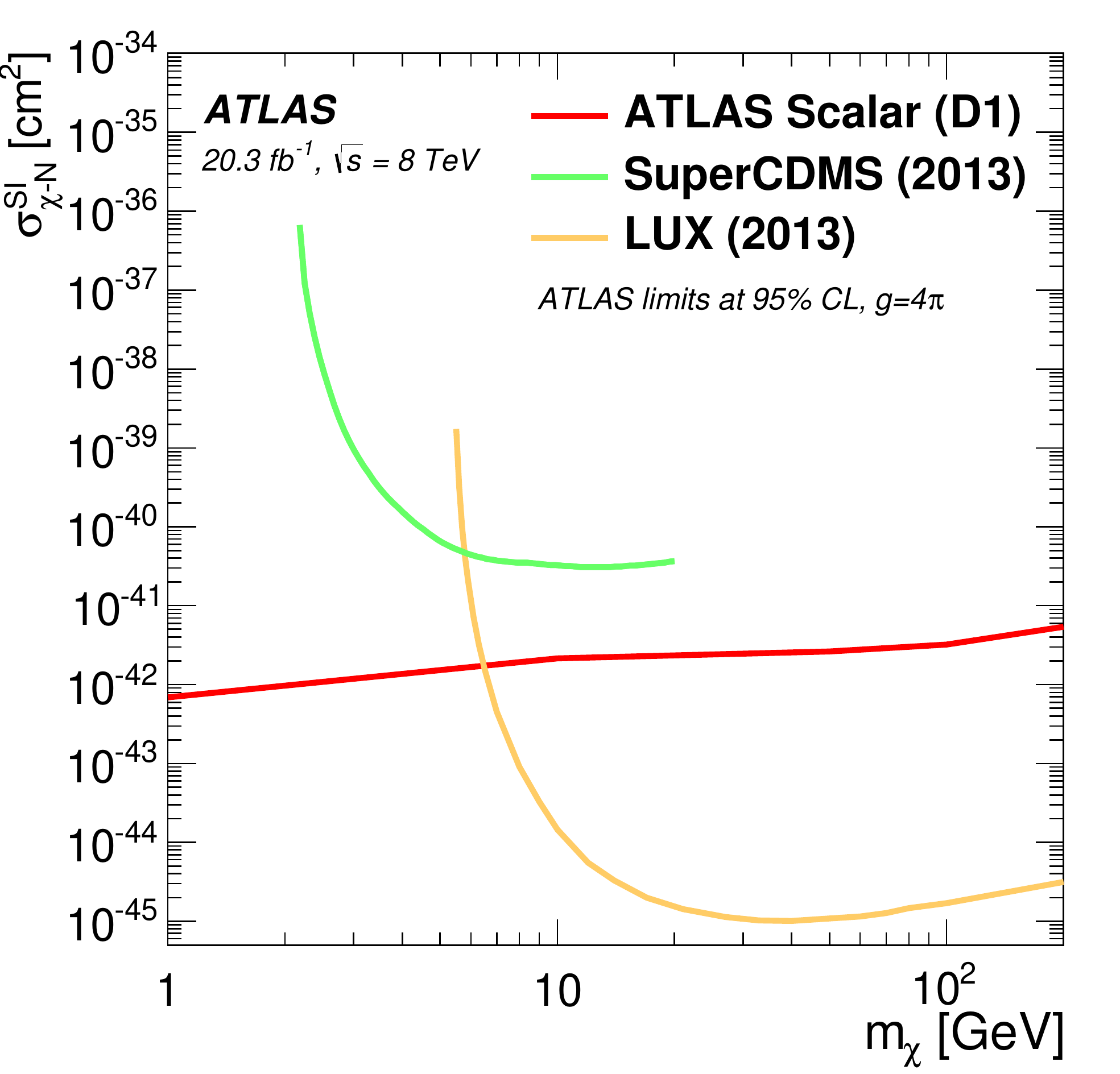}
  \end{tabular} 
  \caption{Spin-independent 95\% CL  upper limits on the DM-Nucleon cross section ($\sigma_\textrm{\tiny $\chi$-N}^{\rm SI}$) 
  for the scalar operator D1 (red)
  as a function of $m_{\chi}$. The yellow and green curves represent the 90\% CL exclusion limits set by the LUX and Super-CDMS collaborations. 
 The coupling is assumed to be $g_q g_\chi=g=4 \pi$.
} 
  \label{figxsection_si_95}
\end{figure}

\begin{figure}[h]
  \centering \begin{tabular}{c@{\hskip 5mm}c}
    \includegraphics[width=.5\textwidth]{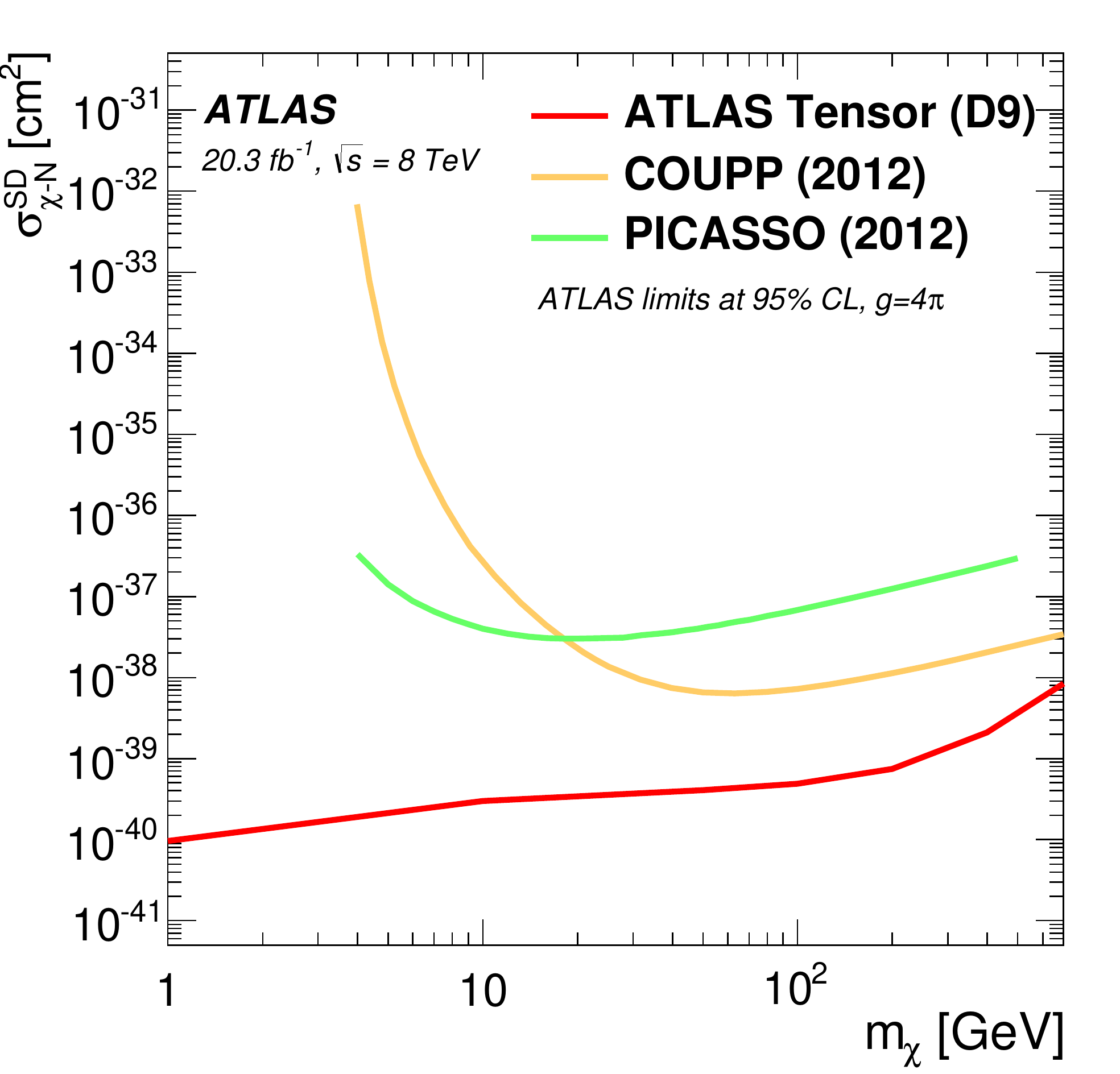}
  \end{tabular} 
  \caption{
Spin-dependent 95\% CL upper limits on the DM-Nucleon cross section ($\sigma_\textrm{\tiny $\chi$-N}^{\rm SD}$) 
for the tensor operator D9 (red)
  as a function of $m_{\chi}$. The yellow and green curves represent the 90\% CL exclusion limits set by the COUPP and PICASSO collaborations. 
 The coupling is assumed to be $g_q g_\chi=g=4 \pi$.
} 
  \label{figxsection_sd_95}
\end{figure}

%% file: atlas_authlist.tex
\begin{flushleft}
{\Large The ATLAS Collaboration}

\bigskip

G.~Aad$^{\rm 85}$,
B.~Abbott$^{\rm 113}$,
J.~Abdallah$^{\rm 152}$,
S.~Abdel~Khalek$^{\rm 117}$,
O.~Abdinov$^{\rm 11}$,
R.~Aben$^{\rm 107}$,
B.~Abi$^{\rm 114}$,
M.~Abolins$^{\rm 90}$,
O.S.~AbouZeid$^{\rm 159}$,
H.~Abramowicz$^{\rm 154}$
% ...
\bigskip
\\
% ...
$^{11}$ Institute of Physics, Azerbaijan Academy of Sciences, Baku, Azerbaijan\\
% ...
$^{85}$ CPPM, Aix-Marseille Universit{\'e} and CNRS/IN2P3, Marseille, France\\
% ...
$^{90}$ Department of Physics and Astronomy, Michigan State University, East Lansing MI, United States of America\\
% ...
$^{107}$ Nikhef National Institute for Subatomic Physics and University of Amsterdam, Amsterdam, Netherlands\\
% ...
$^{113}$ Homer L. Dodge Department of Physics and Astronomy, University of Oklahoma, Norman OK, United States of America\\
$^{114}$ Department of Physics, Oklahoma State University, Stillwater OK, United States of America\\
% ...
$^{117}$ LAL, Universit{\'e} Paris-Sud and CNRS/IN2P3, Orsay, France\\
% ...
$^{152}$ Institute of Physics, Academia Sinica, Taipei, Taiwan\\
% ...
$^{154}$ Raymond and Beverly Sackler School of Physics and Astronomy, Tel Aviv University, Tel Aviv, Israel\\
% ...
$^{159}$ Department of Physics, University of Toronto, Toronto ON, Canada\\
% ...
\end{flushleft}
